\documentclass{eas}
\usepackage{graphicx}

\usepackage{hyperref}
\usepackage{natbib}

\usepackage{array}
\usepackage{amsmath,amssymb}
\usepackage{color}
\usepackage{listings}
\definecolor{codegray}{rgb}{0.95,0.95,0.95}
\newcommand{\mnras}[1]{\emph{Monthly Notices of the Royal Astronomical Society}}
\newcommand{\apj}[1]{\emph{The Astrophysical Journal}}
\newcommand{\apjl}[1]{\emph{The Astrophysical Journal Letters}}
\newcommand{\apjs}[1]{\emph{The Astrophysical Journal Supplement Series}}
\newcommand{\aj}[1]{\emph{The Astronmical Journal}}
\newcommand{\aap}[1]{\emph{Astronomy \& Astrophysics}}
\newcommand{\aaps}[1]{\emph{Astronomy \& Astrophysics Supplement Series}}
\newcommand{\araa}[1]{\emph{Annual Review Astrononmy \& Astrophysics}}

\TitreGlobal{Statistics for Astrophysics: Clustering and Classification}
\Editors{D. Fraix-Burnet \& S. Girard}  
\idline{EAS Publications Series, Vol. 77, \the\year}
\doi{http://dx.doi.org/10.1051/eas/1677010}

\begin{document}
\setcounter{page}{221}

\title{Concepts of Classification and Taxonomy \\ Phylogenetic Classification}
\runningtitle{Phylogenetic Classification}

\author{Didier Fraix-Burnet}
\address{Univ. Grenoble Alpes, IPAG, F-38000 Grenoble, France}
\secondaddress{CNRS, IPAG, F-38000 Grenoble, France}

\begin{abstract}
Phylogenetic approaches to classification have been heavily developed in biology by bioinformaticians. But these techniques have applications in other fields, in particular in linguistics. Their main characteristics is to search for relationships between the objects or species in study, instead of grouping them by similarity. They are thus rather well suited for any kind of evolutionary objects. For nearly fifteen years, astrocladistics has explored the use of Maximum Parsimony (or cladistics) for astronomical objects like galaxies or globular clusters. In this lesson we will learn how it works.
\end{abstract}

\maketitle

\section{Why phylogenetic tools in astrophysics?}
 \subsection{History of classification}
 
The need for classifying living organisms is very ancient, and the first classification system can be dated back to the Greeks. The goal was very practical since it was intended to distinguish between eatable and toxic aliments, or kind and dangerous animals. Simple resemblance was used and has been used for centuries. Basically, until the XVIIIth century, every naturalist chose his own criterion to build a classification. At the end, hundreds of classifications were available, most often incompatible to each other. The criteria for this traditional way of classifying is the subjective appearance of the living organisms.

During the XVIIIth a revolution occurred. Scientists like Adanson and Linn\'e devised new ways of classifying the objects and naming the classes. Adanson realised that all the observable traits should be used, giving birth to the multivariate clustering and classification activity \citep{adanson}. Linn\'e based his binomial nomenclature on neutral names unrelated whatsoever to any property of the classes. We can realise the success of these two ideas more than two centuries and a half later!

The hierarchical organization of the living organisms was already established when Linn\'e devised his nomenclature, but its origin became understood in the mid-XIXth thanks to Darwin. Evolution was the key to understand the hierarchy and interpret the scheme as depicting the evolutionary relationships between the species. Then, biologists have since devoted themselves to establish the phylogenetic tree of all living organisms, the Tree of Life.

Partitioning or hierarchical clustering techniques were used in this purpose by comparing species by their global similarity. However, William Hennig in 1950 translated the transmission with modification idea of Darwin into a new concept of classification: species are not compared anymore on the ressemblance, but on their ascendance. We are not looking anymore for who is like who, but who is cousin of whom. Cladistics was born \citep{hennig1965}. 

Indeed, it seems that linguists already used a similar approach to compare the languages and their diversification, but never formalised it like Hennig. Harsh debates occurred in the evolutionary biology community, but this approach, better called now Maximum Parsimony, was definitively adopted in the 1980s. 

Despite that other techniques were developed and are very often used nowadays, the philosophy of cladistics is still behind all these approaches. Cladistics is conceptually simple and very general, yet it is also very demanding computationally. 

Table~\ref{tab:history} summarizes the evolution of the concepts of classification. It is important to keep this in mind. In astrophysics, we are still essentially using the traditional way of classifying objects.

\begin{table}[h!]  
 \renewcommand{\arraystretch}{1.5}
  \begin{center}
\begin{tabular}{|r|l|l|l|}
\hline
$\sim -300$ & \textbf{Aristote} & First classification & \emph{Appearance, usefulness} \\
\hline
$\sim 1730$ & \textbf{Linn\'e} & Binomial nomenclature & \emph{Objective taxonomy} \\
\hline
1763 & \textbf{Adanson} & Use of all parameters & \emph{Global similarity} \\
\hline
1850 & \textbf{Darwin} & Evolution & \emph{Inheritance, mutations} \\
\hline
1950 & \textbf{Hennig} & Cladistics & \emph{Common history} \\
\hline
  \end{tabular}
  \end{center}
  \caption{\label{tab:history}A summary of the evolution of the concepts of classification.}
  \end{table}

Since astronomical objects are evolving, it is natural to consider that phylogenetic tools might be adapted. The transmission with modification notion is not restricted to the living organisms. Two spiral galaxies that merge create a new object, possibly a galaxy of elliptical shape, which contains the material from its progenitors (transmission) but with modified properties like the kinematics or new generations of stars (modification). It can also be purely virtual. For instance, if you want to transform a red triangle to a red circle, you have to transmit the color but modify the shape.

 \subsection{Two routes for classification}

\begin{figure}[ht]
 \begin{center}
\includegraphics[width=0.6\textwidth]{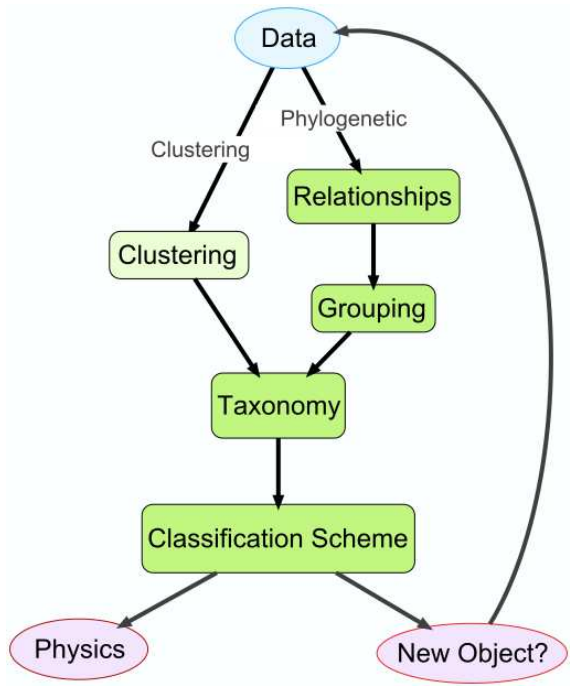}
 \end{center}
 \caption{\label{fig:systematics}The two routes to build a classification scheme. The clustering path gathers objects together, any relationships between the classes are established afterwards using physical or biological arguments. Conversely, the phylogenetic approach looks for the relationships from which groups are defined.}
 \end{figure}

Two thousands years of classification in biology have taught us that there are two paths toward devising a classification scheme (Fig.~\ref{fig:systematics}). 

 The first path is to gather objects according to appearance or global similarity. This is clustering. The groups are then characterized, described and defined, and are given names respecting rigorous taxonomic rules. All this provides the classification scheme that will later be used for (supervised) classification. 
 
 Having grouped objects together is the end of the statistical investigation. Now we wish to understand why the objects are grouped in this way, and what are the relationships between the groups. In other words, what is the origin of the groups? 
 
 In biology, the key is evolution and Darwin showed that the transmission with modification mechanism creates the hierarchical organization of the living organisms, the diversity of which can thus be well represented on a tree-like diagram.
 
 In astrophysics, Hubble made a clustering of the galaxies he discovered, defining categories like elliptical or spiral, and lated hypothesized that the dissipation would flatten elliptical galaxies into disky ones like spiral galaxies. This is the famous Hubble Tuning Fork diagram that established the relationships between the classes after the clustering process. 
 
 Since we are interested in the relationships between species, why not try to gather the living organisms according to their relatedness? This is the idea of the phylogenetic path toward a classification. It establishes the relationships first, and then the grouping of the objects is deduced from the tree.
 
 The phylogenetic approach is the subject of this lecture. In Sect.~\ref{distance} we present a few techniques based on the pairwise distances between the objects in study. In Sect.~\ref{character} we present the techniques that use the parameters themselves, and detail the Maximum Parsimony (or cladistics).

 \subsection{Taxonomy}

Taxonomy is the science of naming, defining, describing and classifying that have been developed for the living organisms. 

It is not simple to cluster objects into classes, and it is not obvious to classify new objects into known classes. This depends very much on the description of the classes. In particular, the names of the classes can have a strong influence on this process. 
   
This difficulty has been recognised by Linn\'e in the XVIIIth century. The enormous success of the nomenclature he proposed is due to the fact that the name of a class is not related to any of its specific properties. Why? Let us take an example: what do you think the class of ``elliptical galaxies'' is made of? Star-forming galaxies? Radio galaxies? Massive galaxies? No, of course, all galaxies with an elliptical shape belong to this class: the name of the class inevitably plays also the role of its definition and its description, whatever is the diversity in the other properties.

This rule is indeed quite general. I give below a few examples of some classes of objects with their name and (short) definition. Try to deduce the kind of objects with the name only:
\begin{itemize}
   \item Bird: class \textit{Aves}, a group of endothermic vertebrates, characterised by feathers, a beak with no teeth, the laying of hard-shelled eggs, a high metabolic rate, a four-chambered heart, and a lightweight but strong skeleton.
   \item Mammals: class \textit{Mammalia}, any members of a clade of endothermic amniotes distinguished from reptiles and birds by the possession of hair, three middle ear bones, mammary glands, and a neocortex (a region of the brain).
   \item Particle: a minute fragment or quantity of matter. Classified as electron, proton, neutron, Z...                                                                                                                           
   \item Star: a luminous sphere of plasma held together by its own gravity. Classified as O A B..., red giant, dwarf...                                                                                                                                         
   \item Galaxy: a gravitationally bound system of stars, stellar remnants, interstellar gas and dust, and dark matter. Classified as spiral, elliptical, ultra-luminous infrared, blue, giant....                                                                                                                                                                                                                   
\end{itemize}
 
 In biology, many species have a common name which is not the scientific name. Note that the name of the class Mammalia strongly suggests the presence of mammary glands... Interestingly, in particle physics, it is impossible to guess the nature of the particles from their names. The quarks have poetic names that has nothing to do with their properties (charm, beauty...).
 
 In astrophysics, this is not so rigorous. The first and most used classification of stars respects the fundamental taxonomic rule, but other classes have been added to this picture following apparent physical or chemical properties, and mainly describing evolutionary stages.
 
 For galaxies, this is far worse, there are hundreds of classifications according to the many observables we managed to obtain. I cannot refrain from seeing here a parallel with the situation of biological classifiation at the end of the XVIIth century that motivated the works by people like Linn\'e and Adanson.
 
 So the requirement of multivariate classifications in astrophysics appears not only natural, but somewhat urgent.
 
\subsection{Stars, galaxies: multivariate objects in evolution}

 Evolution is ubiquitous in our expanding Universe, and a multivariate description of astrophysical objects is necessary to distinguish different species, to relate together the evolutionary stages or the progenitors, and to build and constrain our physical models. 
 
When two spiral (disky) galaxies merge, they generally yield a galaxy with an elliptical shape. When a huge cloud of gas collapses, it can also yield an elliptical galaxies. These two scenarios, validated by numerical simulations, implies extremely different evolutionary histories for the evolution and formation of galaxies. It has also great consequences for our knowledge of the evolution of our Universe: in the first case a rather long time is necessary, while in the second case it can happen rather fast and early. Yet, the final product looks the same when considering only the morphology. 

It is quite instructive to recognise these two different populations of elliptical galaxies. For this purpose, other properties like the kinematics and the gas content should be of good help. 

This little example is a convincing case for multivariate and evolutionary studies of the diversity of astrophysical entities. This is precisely the goal of phylogenetic analyses \citep[e.g. see a review of multivariate approaches for galaxy classification in ][]{TF15}.

\section{Distance-based approaches}
\label{distance}
 
 \subsection{Minimum Spanning Tree (MST)}

  A spanning tree is an acyclic, connected graph $G = (V,E)$, where $V$ are vertices (nodes) and $E$ edges (branches) plus a weight function:
 \begin{align}
  w : &E \rightarrow \mathbb{R}  \\ 
  &e \rightarrow w(e) 
 \end{align}
The Minimum Spanning Tree is the tree $T$ minimizing: 
  \begin{align}  
w(T) = \sum_{e\ \in\ T} w(e) 
\end{align}

  If the weights $w(e)$ are distinct, then the solution is unique.
  
  To perform this computation, we need a matrix with the objects, that will be the nodes, and the weights between any two of them. These weights can be pairwise distances computed from the parameters describing the objects, or dissimarities, or any other ``cost'' that is pertinent for the study.
    
How can we make a clustering from a MST? The branches depict some relationships between the connected points according to the minimisation.
By cutting the longest branches, the ones with $w(e) > w_{\mathrm{threshold}}$, we can separate bunches of branches, effectively creating groups.  This is sometimes called the separation of the tree. This process is similar to the simple linkage criterion in the hierarchical clustering method, and the friends-of-friends algorithm or Nearest Neighbor algorithm known to astronomers \citep{Gower1969,FeigelsonBabu2012}.

The MST technique has been long used by cosmologists to map the cosmic web and spatially clustering galaxies \citep{Barrow1985}. To get a better idea of the distinct filaments and their respective orientations, it is possible to prune the tree, i.e. to remove some branches, in particular the ones which are leading to a single terminal node (degree $k$ = 1) and originate from a node of degree $k$=3 (three branches connects to this node).

\begin{figure}[ht]
\begin{center}
 \includegraphics[width=0.5\textwidth]{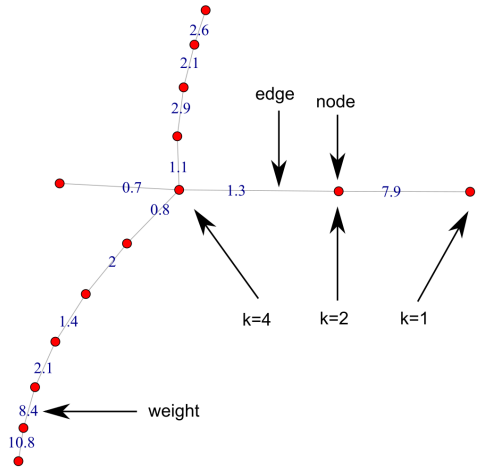}
\caption{\label{fig:MST}An example of a Minimum Spanning Tree with some definitions (see text). The index $k$ is the degree of a node, i.e. the number of branches connected to it.}
\end{center}
\end{figure}

 \subsection{Neighbor Joining Tree Estimation (NJ)}

\begin{figure}
\begin{center}
  \includegraphics[width=0.6\textwidth]{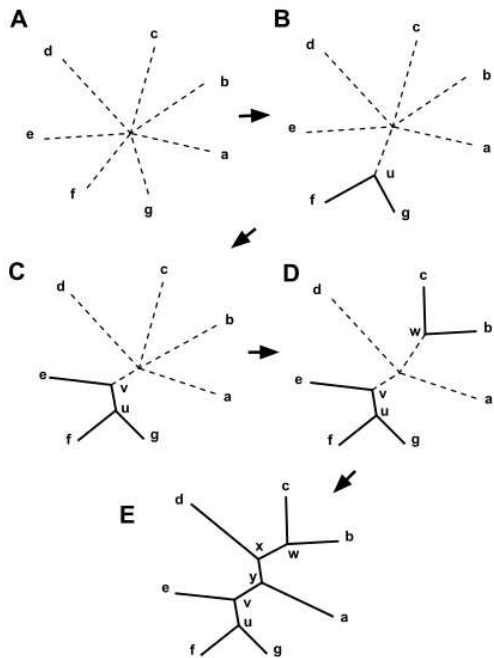}                                                                                                      
  \caption{\label{fig:NJ}A diagram to show the successive steps of a Neighbor-Joining Tree Estimation (Source: 
\url{https://en.wikipedia.org/wiki/Neighbor_joining}).}
\end{center}
  \end{figure}

 Neighbor-Joining \citep[NJ,][]{NJ1987,NJ2006} is the most popular distance-based approach to construct a phylogenetic tree. This is a bottom-up hierarchical clustering methods starting from a star tree (unresolved tree). The procedure runs as follows (see Fig.~\ref{fig:NJ}):
 
\textbf{1.} from the distance (or dissimilarity) matrix for the $n$ objects giving the distances $d(i,j)$ between objects $i$ and $j$, compute the ``corrected'' distance $Q(i,j)$ which represents a kind of evolutionary distance measured on the tree:

\begin{equation}
  Q(i,j) =  (n-2)d(i,j) - \sum_{k=1}^{n}d(i,k)  - \sum_{k=1}^{n}d(j,k)   
\end{equation}
                                                                           
\textbf{2.} find the lowest $Q$ and relate the two corresponding objects with a new node $u$. This new node is called an internal node.

\textbf{3.} replace $i$ and $j$ by $u$, rebuild the new distance matrix with the $n-1$ objects using the following definition:
  
  \begin{equation}
     d(u,k) = \frac{1}{2}\left[d(i,k)-d(i,u)\right] +  \frac{1}{2}\left[d(j,k)-d(j,u)\right] 
  \end{equation}

  \textbf{4.}  iterate from 1 until all the objects have been considered.
  
  Neighbor-Joining minimizes a tree length, according to a criteria that can be viewed as a Balanced Minimum Evolution \citep{NJ2006}. It furnishes a simple algorithm to reconstruct a tree from the distance matrix.

 \subsection{Difference between a hierarchical tree and a phylogenetic tree}
 
 \begin{figure}[ht]
\begin{center}
  \includegraphics[width=0.8\textwidth]{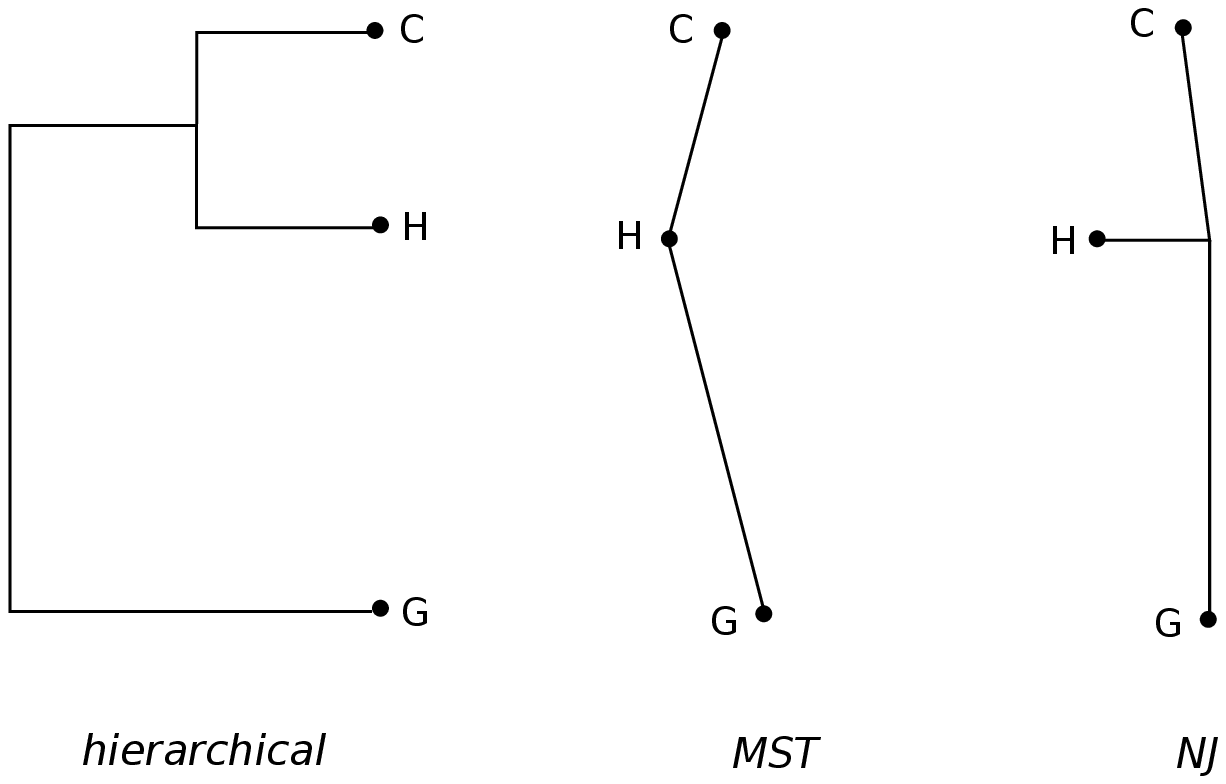}                                                                                                      \end{center}
\caption{\label{fig:hiera_MST_NJ}Three different kinds of trees representating the respective situations of the cities of Chamonix (C), Les Houches (H) and Geneva (G). Left: a hierarchical clustering tree. Middle: a MST tree. Right: A NJ tree showing the addition of an internal node. See text for details.}
 \end{figure}

 The hierarchical clustering trees, the MST and the NJ trees are all based on distances computed from the parameters and are used to define groups. Are there really different and does it matter? Let us take a simple and concrete example (Fig.~\ref{fig:hiera_MST_NJ}).
 
 The hierarchical tree depicts the relative distances between the objects. For instance, it tells you that Les Houches and Chamonix are much closer to each other than they are from Geneva. But this does not tell you whether, when you will leave the conference centre in Les Houches to go the Geneva airport, you should go through Chamonix. A hierarchical clustering tree does not provide you with the relationships between the cities. 
 
 For this you need a phylogenetic tree. The MST tells you that from Les Houches you should go directly to Geneva without visiting Chamonix. The relationships situates Les Houches in between Chamonix and Geneva.
 
 But this is not entirely correct, since the road between Chamonix and Geneva does not go through the village of Les Houches. There is crossroad just outside Les Houches. The reason is that whatever your departure point to go to Geneva, you will take a shorter path. This crossroad is an ``internal node'' introduced in the NJ method, as well as the cladistics we are going to present.

\section{Character-based approaches}
\label{character}

Parameters that, after discretization, can be given ancestral and derived states are called characters. This means that these parameters describing the objects to classify have kept a trace of the historical evolution of the different species. 

A big advantage of character-based approaches is their ability to take uncertainties or unknowns into account. It is of course not possible to compute a distance when a data is missing. Here, one can simply evaluate the different possibilities allowed by a proposed range of values, and then select the best ones according to the optimization criterion. In the case of unknown parameters, the phylogenetic tree thus provides a prediction for the unknown values.

\subsection{Maximum Parsimony (cladistics) and Maximum Likelihood}

 The Maximum Parsimony (cladistics) algorithm compares objects according to their shared common histories using the characters. It selects the phylogenetic tree that minimizes the total number of state changes, which depicts the simplest evolution scenario given the data set \citep{Felsenstein1984}.
 
 Maximum Parsimony is a powerful approach to find tree-like arrangements of objects. Its main drawback is that the analysis must consider all possible trees before selecting the most parsimonious one. The computation complexity depends on the number of objects and character states, so that too large samples (say more than a few thousands) cannot be analyzed.

 Another class of character-based techniques relies on an a priori models of character evolution (probabilites of state changes). For instance, the Maximum Likelihood algorithm selects the most probable phylogenetic tree given the proposed evolutionary model \citep{Williams2003}.
  This kind of approaches are mainly developed for genetic data (evolution of nucleotide sites, mutation and substitution rates of genes...) and they can be applied to big data sets.

In this chapter we present cladistics, or Maximum Parsimony, in great details since it is conceptually the simplest and also the most general phylogenetic method.

\subsection{Cladistics: constructing a Tree}
\label{constructatree}

 Let us begin with a very simple case. Consider three families of objects, E, S, and Sb, characterized by two parameters: Arm and Bar. These parameters can be present (code 1) or absent (code 0). The Table~\ref{tab:simple} gives our data.

 \begin{table}[h!] 
  \renewcommand{\arraystretch}{1.5}
\begin{center}
	\centering
		\begin{tabular}{|c|ccc|}
            \hline
			    &  &  Arm  &  Bar \\
            \hline
			E    &  &  0   &  0  \\
			S    &  &  1   &  0   \\
			Sb   &  &  1   &  1   \\
            \hline
		\end{tabular}
\end{center}
   \caption{\label{tab:simple}Three families E, S and Sb, of objects described with two parameters Arm and Bar which can be present (1) or absent (0).}
\end{table}

There is only one possible tree to link the three objects with a tree. It is given in the graph below (left). 

\begin{figure}[ht]
     \label{fig:simple}
  \begin{center}   
 \includegraphics[width=3 cm]{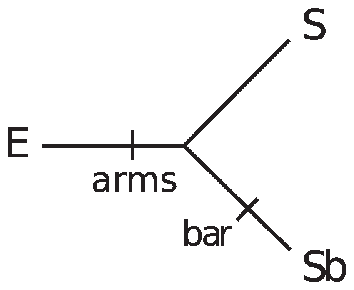} \hfil
    \includegraphics[width=5 cm]{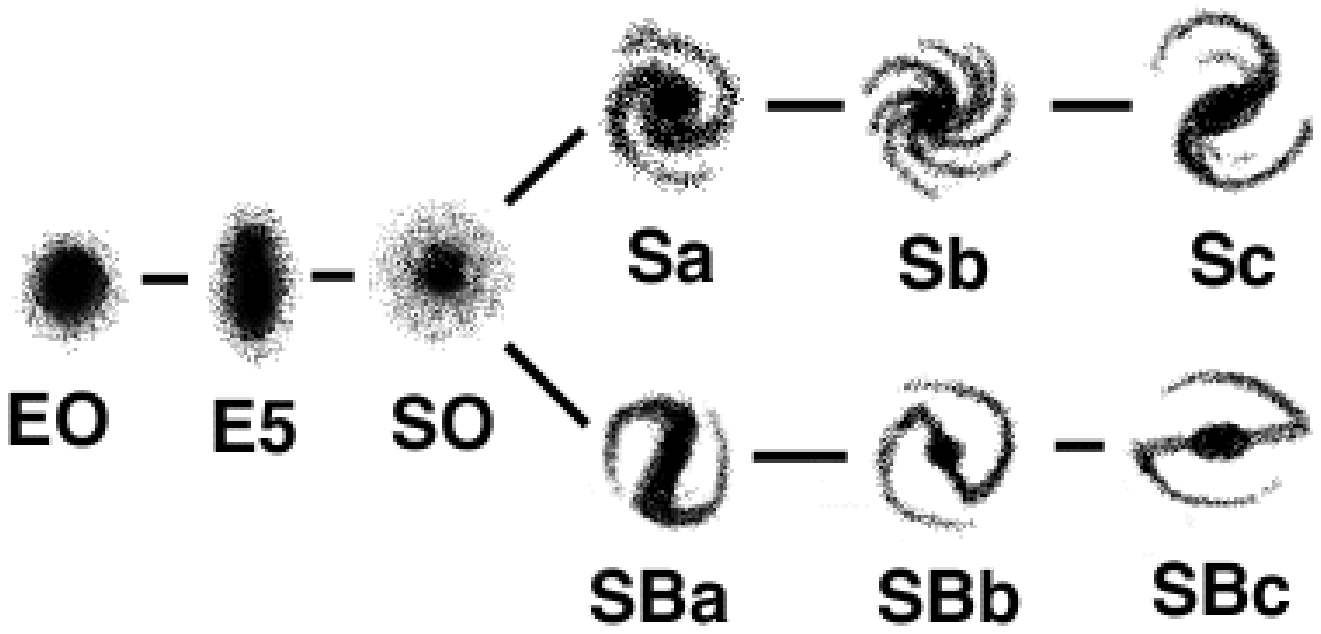} \hfil
    \includegraphics[width=3 cm]{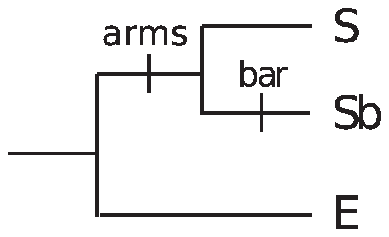}
\end{center}
\caption{Left: the unrooted tree that can be built with three objects (see Table~\ref{tab:simple}). Middle: the famous Hubble Tuning Fork diagram. Right: the rooted tree corresponding to the left figure.}
\end{figure}

If we follow the evolution from E to S, we have to change the Arm parameter from 0 to 1. This is indicated by the tick mark on the branch starting from E. To go from E to Sb, we have to add a tick mark before Sb to indicate a change in the parameter  Bar from 0 to 1. Going from Sb to S means changing the Bar parameter from 1 to 0 as already indicated by the tick mark on the Sb branch. With such a diagram and two tick marks, we have depicted entirely the table above with the possible (but still hypothetical at this stage) evolutionary scenario.

Obviously this sample describes the simplest classification of galaxies, the one using morphology: Elliptical, Spiral and Barred Spiral galaxies. It is striking to note that the cladogram on the left is so much similar to the famous Hubble Tuning Fork diagram (middle) that Hubble himself drew to depict is thought on galaxy evolution in the 1930's! 

With a cladogram, we can go one step further by adding an arrow of evolution (biologists prefer to say diversification). In this purpose, if we impose or know that the state ``0'' is more primitive (ancestral) than the state ``1'' (derived), in other words spiral structures and bars appeared in the course of the evolution of the Universe, we obtain the so-called rooted diagram depicted on the right. Note that the leftmost branch and node are not attributed to any objects. This is a rule, since it is always possible that a new object will be discovered and will have to be inserted somewhere in the tree. Hence, the common ancestor of a group is always assumed to be unknown.

   Consider the case with four objects described by only one parameter as given in Table~\ref{tab:fourand1}.
\begin{table}[h!] 
 \renewcommand{\arraystretch}{1.5}
\begin{center}
	\centering
		\begin{tabular}{|c|cc|}
            \hline
			O   &  &   0    \\
			A   &  &   1    \\
			B   &  &   1    \\
			C   &  &   0    \\
            \hline
		\end{tabular}
\end{center}
   \caption{\label{tab:fourand1}Four objects described by one parameter with two states, 0 and 1.}
 \end{table}
 
You can make the exercice to find all the possible arrangements of the objects on a tree (unrooted). The result is that there are now three possible solutions. This number depends only on the number of objects, and increases incredibly fast with it. It becomes very difficult (and boring) to draw them all by hand for more than 7 or 8 objects. Even modern computers take a huge amount of time to explore the exhaustive tree space for more than a hundred trees.

\begin{figure}[h!]
\begin{center}
  \includegraphics[width=\textwidth]{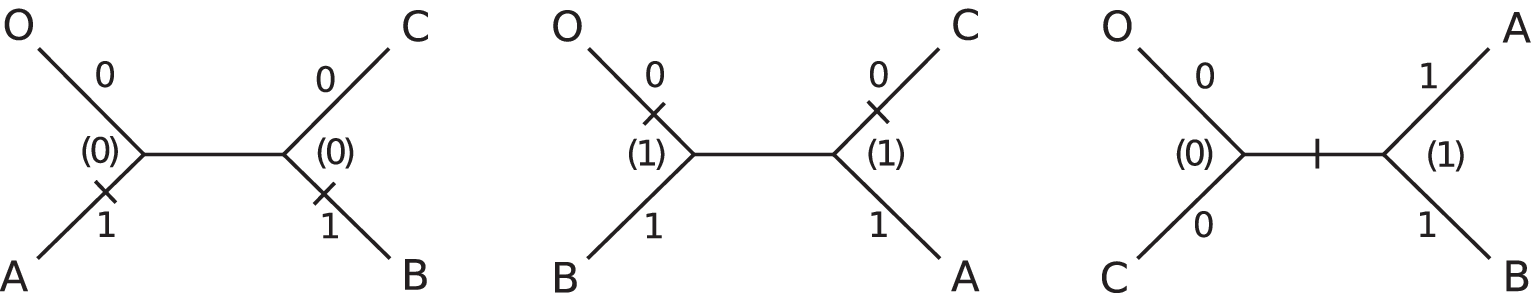}
  \vskip 15pt
  \includegraphics[width=\textwidth]{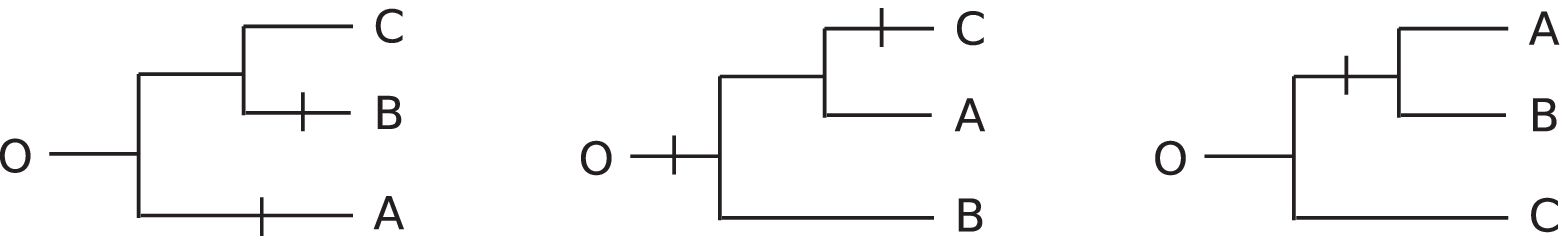}
\end{center}
\caption{\label{fig:fourand1}The three possible arrangements of four objects on a tree (top: unrooted, bottom: rooted). The two trees to the left have a total score of two steps, and the right tree is the simplest scenario with only one step.}
\end{figure}

The three possible arrangements with four objects are shown in Fig.~\ref{fig:fourand1}. For each tree, we have to indicate the parameter changes to evolve from any one of the objects to any other ones, choosing the simplest solution to put as few tick marks as possible.
Each tick mark is called a step, and the total number of steps found on a tree indicates the cost of evolution. This is a measure of the complexity of the evolutionary scenario depicted by the tree. We notice that the tree on the right has only one step, it is the simplest of all the possible arrangements for this four objects, i.e. the most parsimonious tree that is chosen as the most probable evolutionary scenario.

On the unrooted trees, we have added in parentheses the value of the parameter at the node. This is for clarity, and it is not necessary in practice to count the number of steps.

In Fig.~\ref{fig:fourand1}, the rooting of the trees has assumed that the value ``0'' is the ancestral state. As an exercice you can draw the rooted trees assuming the alternative that the value ``1'' is the ancestral value, and count the number of required changes. You will find that the total score of a tree does not depend on the choice of the root.

Consider the sample data given in Table~\ref{tab:fourand5}. We have the same four objects as previously, so that, the possible arrangements on a tree depending only on the number of objects, the  possible trees with four objects are identical to the ones depicted in Fig.~\ref{fig:fourand1}. But we have more information on these objects since they are described by five parameters. The total number of steps depends strongly on the number of parameters. 

\begin{table}[h!]  
 \renewcommand{\arraystretch}{1.5}
\begin{center}
	\centering
		\begin{tabular}{|c|cccccc|}
            \hline
			    &  &  c1  &  c2 &  c3  & c4 & c5 \\
            \hline
			O   &  &  0   &  0  &  0   &  0 & 0 \\
			A   &  &  0   &  0  &  1   &  1 & 1 \\
			B   &  &  0   &  1  &  0   &  0 & 0 \\
			C   &  &  0   &  1  &  1   &  0 & 1 \\
            \hline
		\end{tabular}
\end{center}
\caption{\label{tab:fourand5}Four objects described by five binary parameters.}
\end{table}
Note that the parameter c0 is constant, it is non-informative and can be dropped. The unrooted trees are the same as previously, but the rooting depends on the parameter states and the choice of the ancestral states. Assuming ``0'' as the ancestral state for all the parameters, object O is obviously the most ancestral object and we show only the rooted trees.

\begin{figure}[h!]
\begin{center}
    \includegraphics[width=\textwidth]{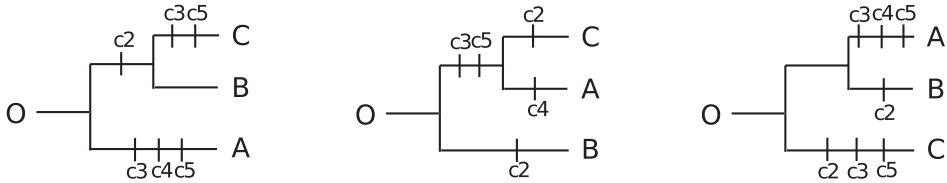}
\end{center}
\caption{\label{fig:fourand5}Trees obtained with the sample data given in Table~\ref{tab:fourand5}. From left to right, the scores of the trees are 6, 5 and 7 steps.}
\end{figure}

As shown in Fig.~\ref{fig:fourand5}, the number of tick marks is necessarily higher than previously (Fig.~\ref{fig:fourand1}), and the simplest tree is now the one in the middle. This illustrates the important point that a classification, or an evolutionary scenario changes depending on the data availability.

Up to now, we have considered parameters with only two states, 0 or 1. The Table~\ref{tab:fourandsev} contains four objects as before, and three parameters with different number of states (up to 5 for parameter c2).

\begin{table}[h!]  
 \renewcommand{\arraystretch}{1.5}
\begin{center}
	\centering
		\begin{tabular}{|c|cccc|}
            \hline
			    &  &  c1  &  c2 &  c3  \\
\hline
			O   &  &  0   &  0  &  0   \\
			A   &  &  1   &  2  &  1   \\
			B   &  &  2   &  0  &  1   \\
			C   &  &  3   &  4  &  2   \\
            \hline
		\end{tabular}
\end{center}
\caption{\label{tab:fourandsev}Four objects described by three parameters with several states.}
\end{table}

\begin{figure}[h!]
\begin{center}
 \includegraphics[width=\textwidth]{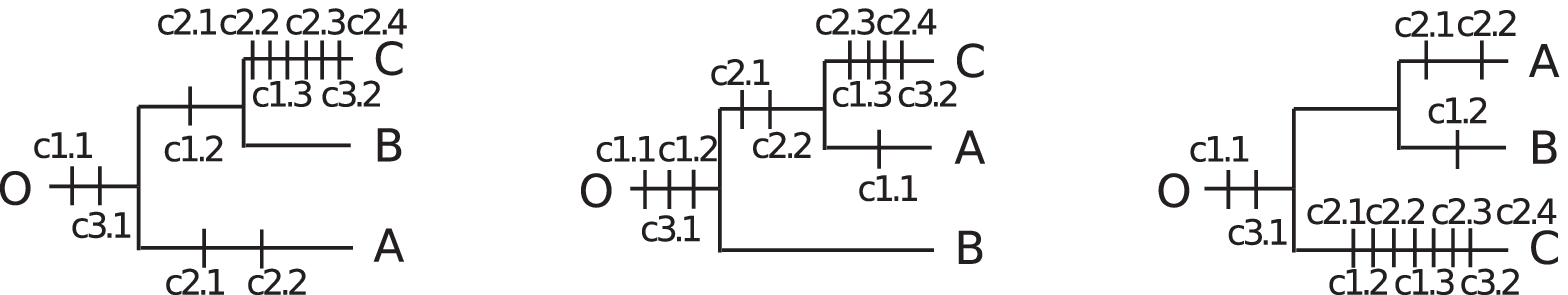}
\end{center}
\caption{\label{fig:fourandsev}Trees obtained with the sample data given in Table~\ref{tab:fourandsev}. From left to right, the scores of the trees are 11, 10 and 12 steps.}
\end{figure}

The unrooted trees are the same, the rooted ones also if we again assume ``0'' as the ancestral state, and thus object O as root, but the tick marks are more complicated to read. In Fig.~\ref{fig:fourandsev} we have adopted the following notation: cn.x means character cn goes from x to x-1 or reversely.

Cladistics is not more complicated than that. Computers come to our help because searching for the simplest tree is rather tedious. Nevertheless, there are some important questions to address before performing analyses with real data.

\subsection{Parameters as Characters}

Cladistics compares objects from innovations inherited from a common ancestor. Characters are parameters that can trace this process of transmission with modification. States are discrete values taken by the characters, they supposedly describe the evolutionary stages of the character. To keep trace of an innovation, characters must have the right evolutionary behaviour.

\smallskip

\begin{minipage}[c]{0.5\linewidth}\textbf{Synapomorphy}
 
A state is specific to a clade (an ancestor and all its descendants). On the example to the right, the value ``disk'' of the character ``shape'' defines a clade gathering objects that inherited this property from a common ancestor. The five objects derive from a common ancestor, and thus define a clade, but this should be justified by other parameters than the shape. 

\end{minipage}
\begin{minipage}[c]{0.5\linewidth}\begin{center}
  \framebox{\includegraphics[width=0.5\textwidth]{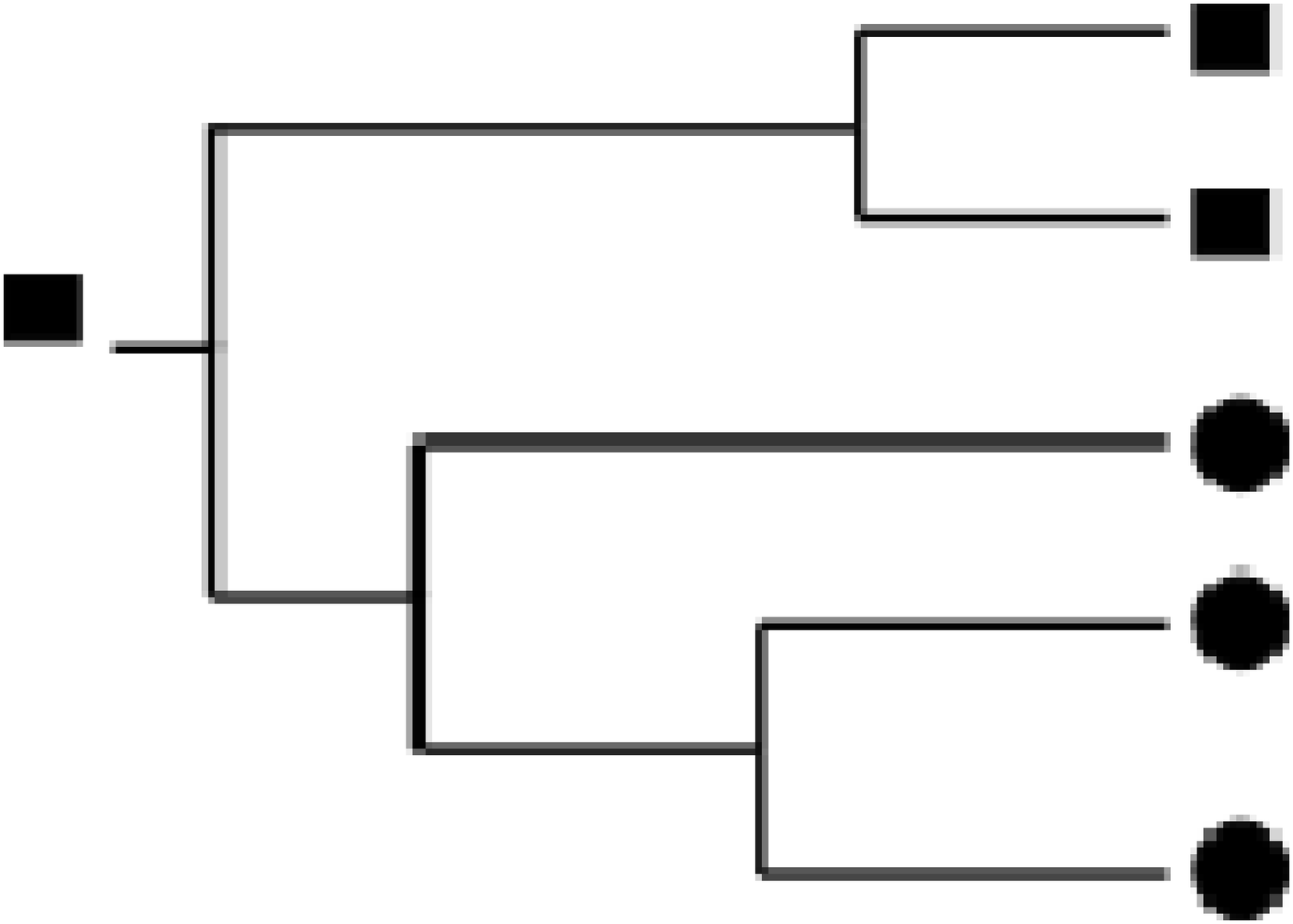}   }  
\end{center}
\end{minipage}

\bigskip

Synapomorphies are the ideal situation for a phylogenetic reconstruction. The following behaviours (called homoplasies) must be avoided because they can destroy a tree:

\medskip

\begin{minipage}[c]{0.5\linewidth}\textbf{Reversal}

An innovative state is lost again in favour of a previous character state. Here, the disk value is lost and the character evolves back to the square shape.
\end{minipage}
\begin{minipage}[c]{0.5\linewidth}\begin{center}  \framebox{\includegraphics[width=0.5\textwidth]{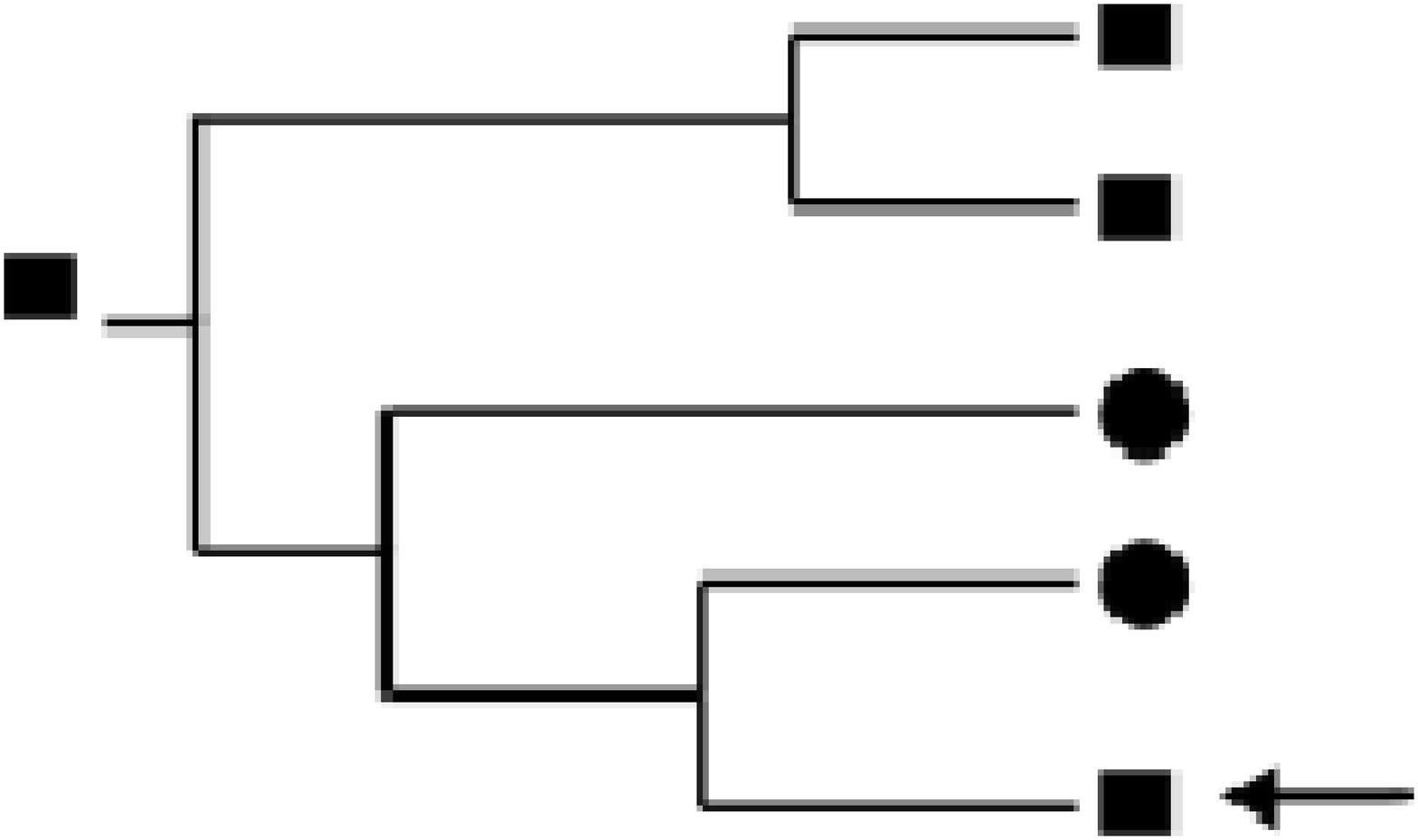}} \end{center}
\end{minipage}

\smallskip

\begin{minipage}[c]{0.5\linewidth}\textbf{Parallel Evolution}

The same evolutionary sequence occurs on two different paths, like the square to disk state evolution here.
\end{minipage}
\begin{minipage}[c]{0.5\linewidth}\begin{center} \framebox{\includegraphics[width=0.5\textwidth]{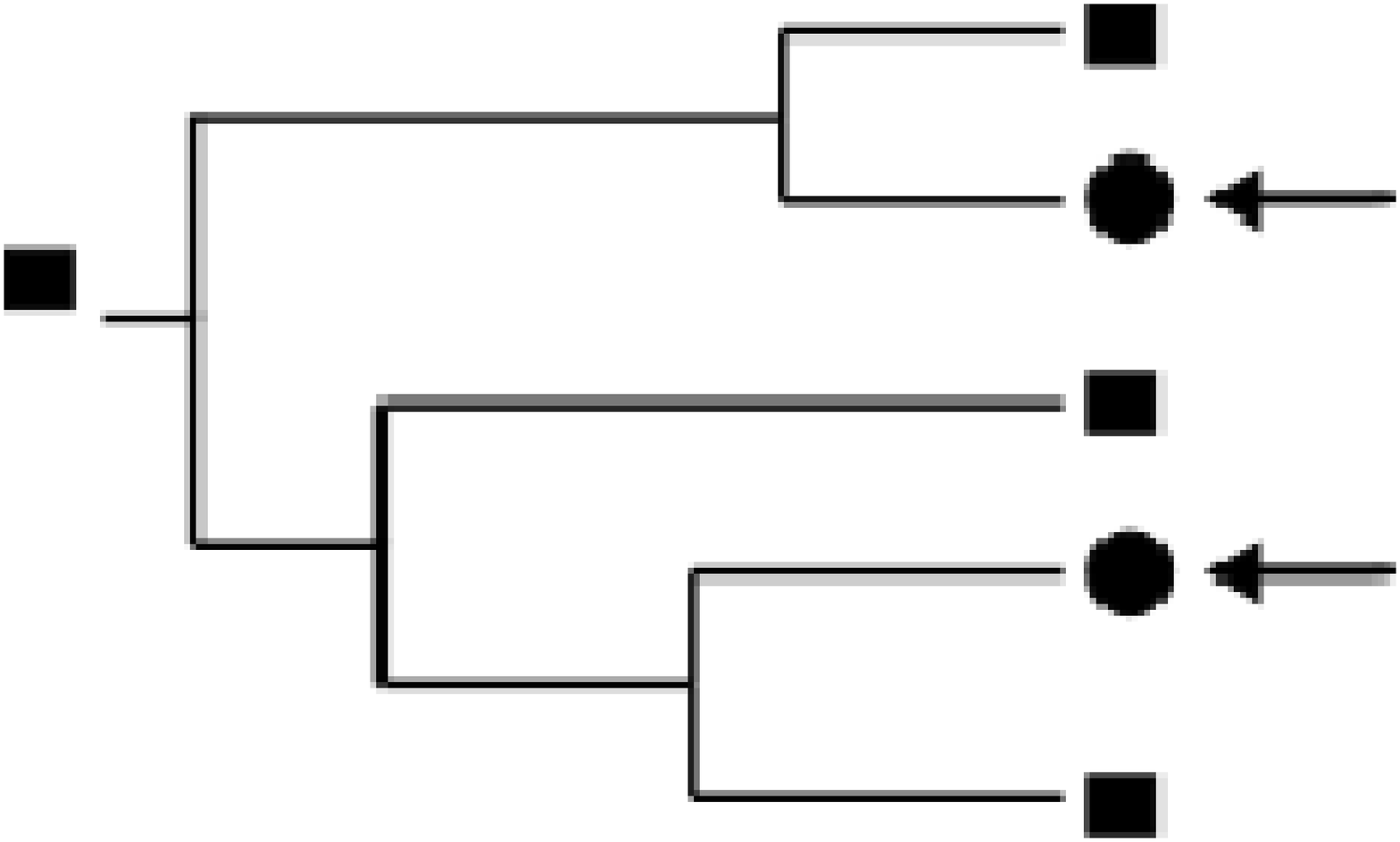}} \end{center}
\end{minipage}

\smallskip

\begin{minipage}[c]{0.5\linewidth}\textbf{Convergence}
   
The same character state results from different evolutionary sequences. Here, the disk value has occured both from square and from star states.
\end{minipage}
\begin{minipage}[c]{0.5\linewidth}\begin{center} \framebox{\includegraphics[width=0.5\textwidth]{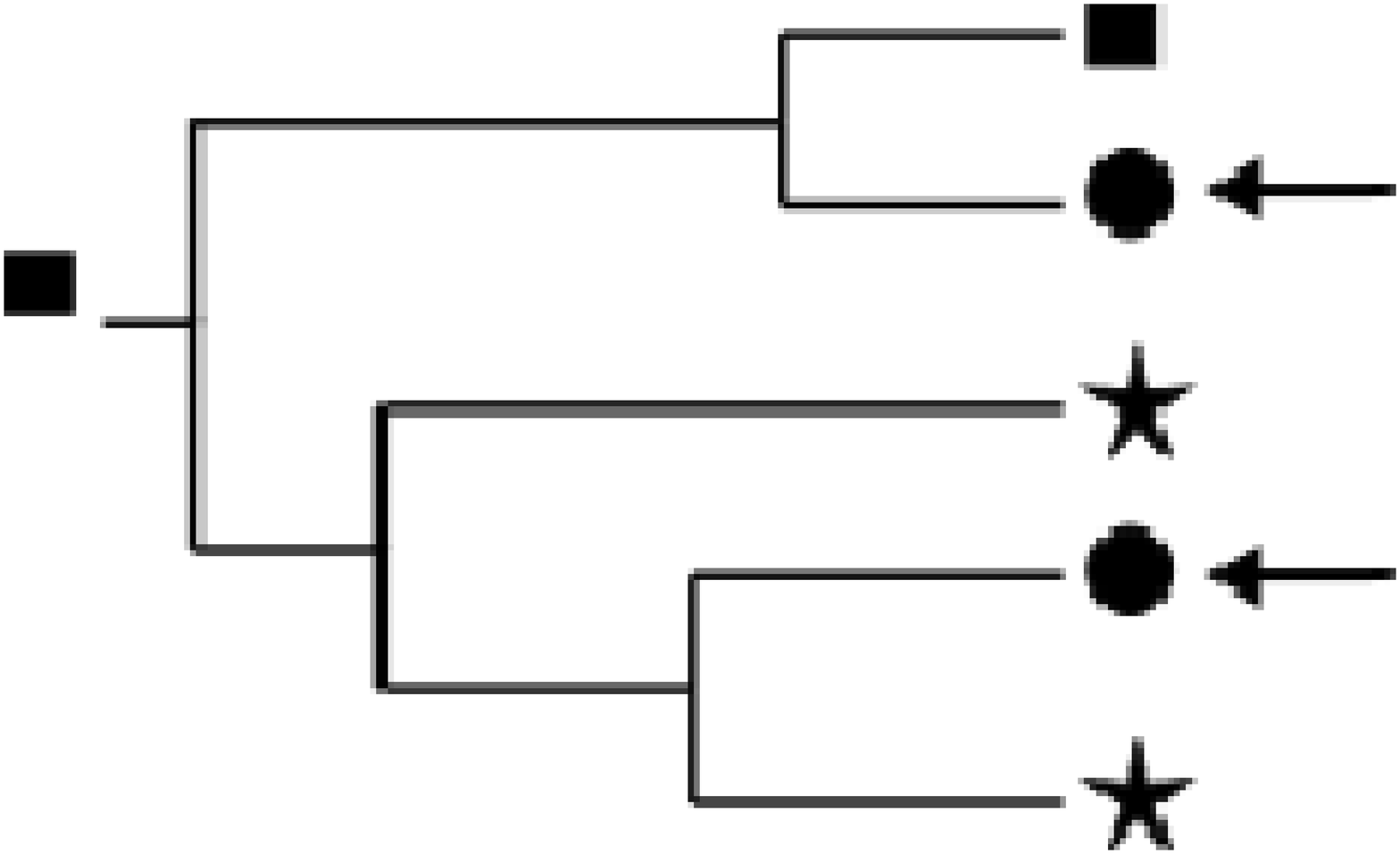}           }                         
                                  \end{center}
\end{minipage}

\bigskip
Some homoplasic characters are allowed if they are not too many. In practice it is difficult to avoid them. Most often the behaviours of characters can be determined once the phylogeny is established, so that there is no other way to find the synapomorphies than a trials and errors approach.

The Maximum Parsimony search minimizes the number of homoplasies and maximizes the number of synapomorphies.

 \subsection{Counting the steps: the cost of evolution}
To evaluate the complexity of trees, we count the number of steps. Up to now, we have assumed a very simple algebra, the difference between two consecutive integer states being one step. However, is it possible to define more complex (realistic?) character transformations. Here are some common hypotheses ($i$ and $j$ are states of character $k$, $f_k(i)$ being the value of state $i$. Usually  $f_k(i)=i$):
\begin{itemize}
   \item \textbf{Ordered or Wagner}: states are reversible and additive: $d=\left|f_k(i) - f_k(j)\right|$.
   \item \textbf{Unordered or Fitch}: states are reversible and non-additive: $d=1$ if $i\neq j$, $0$ if $i=j$.
   \item \textbf{Dollo}: each state occurs only once: no reversal, no parallel evolution.
   \item \textbf{Camin-Sokal}: states are irreversible: ($\forall i>j, f_k(i) > f_k(j)$) or ($\forall i>j, f_k(i) < f_k(j)$).
\end{itemize}

\subsection{The most Parsimonious Tree}

The score of a tree is the total number of steps (change of parameters values of states). This is the total cost of evolution $S = \sum_k\sum_{i>j} d\left(f_k(i),f_k(j)\right)$, given the choice of $d$ made previously.

 Among all possible trees, we choose the one with the lowest score, called the most parsimonious tree (Ockham's razor). This is why cladistics is also called the Maximum Parsimony method.

\begin{figure}[ht]
 \begin{center}
\includegraphics[width=0.4\textwidth,height=0.8\textwidth,angle=90]{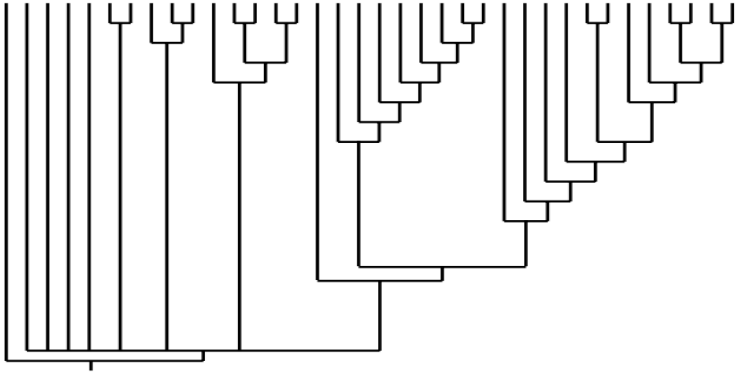}                                                                         
 \end{center}
\caption{\label{fig:constree}Strict consensus tree (100\% nodes only).}
\end{figure}

It is frequent that several equally most parsimonious trees are found. To synthetize the result, we build consensus Tree. Each node of this tree is valued with the percentage of occurence of this node among all the most parsimonious trees. A example of a strict consensus tree (100\% nodes only) is shown in Fig.~\ref{fig:constree}.

Bifurcating nodes (from which three branches emerge) are robust nodes since they occur in all possible most parsimonious trees. They are said to be resolved nodes. Multifurcating nodes indicate a lack of information or constraint from the data. This may also be due to too many homoplasic characters. It is commonly assumed that nodes with a consensus percentage higher than 70\% are quite safe, and lower 50\% not significant at all. Note that the consensus of two perfectly resolved but different trees is not perfectly resolved, meaning that there are two possible robust solutions. Only the interpretation can decide which is the most interesting.

It might be interesting at this level to compare the MST and MP techniques since they both search for a minimum path (Table~\ref{tab:MSTvsMP}).

\begin{table}[ht!]
 \renewcommand{\arraystretch}{1.2}
\begin{center}
   \begin{tabular}{|c|c|}
   \hline
      \textbf{MST} & \textbf{Cladistics} \\
      \hline
&\\
Labelled tree & Semi-labelled tree \\
No internal node & More general \\
Polynomial time & NP-hard \\
$w(T) = \sum_e w(e)$&$ S = \sum_k\sum_{i>j} d\left(f_k(i),f_k(j)\right)$\\
&\\
\hline
\multicolumn{2}{|c|}{For a l1 norm:}\\
&\\
$w(e)=\sum_k\left| f_k(i)-f_k(j) \right|$&$d\left(f_k(i),f_k(j)\right)=\sum_k\left| f_k(i)-f_k(j) \right|$\\
&\\
$w(T)=\sum_{i>j}\sum_k\left| f_k(i)-f_k(j) \right|$&$S= \sum_k\sum_{i>j}\left| f_k(i)-f_k(j) \right|$\\
&\\
\hline
   \end{tabular}
   \end{center}
   \caption{\label{tab:MSTvsMP}Comparison between the MST and the MP techniques.}
   \end{table}

\bigskip

Interestingly, the score of a MP tree is bounded by the total weight of the MST tree:
   $$\frac{w_{min}}{2} < S_{min} < w_{min}$$
   
   The two techniques are thus rather similar, especially for the l1 norm (Manhattan distance), the main differences being the supplementary internal nodes in the MP approach. This implies that MP is more general and can depict more complicated relationships.

\subsection{Robustness Assessment}

Basically, the assessment of the reliability of a tree is based on consensus trees. There is no rigorous mathematical tool, but rather some practical approaches:



\begin{itemize}
   \item \textbf{Trials and errors}: Number of most parsimonious trees, consensus tree resolution, analyses with subsets of characters and objects.... and the interpretation.
   \item \textbf{Bootstrap}: Random draw of characters with replacement.
   \item \textbf{Decay or Bremer degeneracy index}: Number of supplementary steps necessary to have the node disappear.
   \item \textbf{Characters indices}: Retention Index (RI), Consistency Index (CI), Rescaled CI (RCI) measure how much each character supports the tree structure.
\end{itemize}

\subsection{Continuous Parameters}

\subsubsection{Discretization}

In cladistics, we count the scores of trees using the evolutionary states of the characters. How can we deal with continuous parameters? 
   
 Unfortunately, no algorithm has been developed to perform Maximum Parsimony in this case. This is probably because the complexity of the search depends on the number of the character states which is essentially infinite in the continuous case. We must however mention the software program TNT \citep{TNT2008} that allows for 65~000 states, but it assumes some evolutionary model for the character changes that may not be appropriate for astrophysical objects. 

The only solution, often used in biology, and the one that has been used up to now in astrocladistics, is to discretize the continuous parameters. Yes but how? This is a long story and a very difficult question. The simplest way is to divide each character into equal bins. But should they be equal? How many bins? These two questions have no answer in the absolute, and the result depends more or less on the choices made at this level. Another typical question in astrophysics is whether to consider the variables themselves or their logarithms. Probably, the best approach is to try several choices and compare.

In astrocladistics so far, between 10 to 32 equal bins have been used to discretize the parameters. At least, the result should not depend on the precise number of bins and I have found that this is the case between 15 and 32. 

Nevertheless, research is going on about this problem. One interesting direction is to optimize the number and the nature of the bins to ensure to get the best and the most robust tree as allowed by the data. For instance, we begin to understand the mathematical relationships between distance-based and character-based approaches: \cite{TF15} have shown that the two approaches are identical if a multistate character can be reduced to a binary one according to a precise rule (the four-gamete rule).

\subsubsection{Continuous Parameter Evolution}
   
   Finding a tree is never guaranteed since it depends so much on the nature of the sample data and the parameters. Selecting parameters or discovering disturbing objects requires many trials and errors computation. Graph theory may bring interesting help, especially in the case of continuous characters. For instance, when the phylogeny is perfect (in the sense that characters are all synapomorphies), some mathematical rules (the Kalmanson inequalities) holds. 
It is thus possible to search for the best order (arrangement on a tree) to be as close as possible to the fulfilment of the Kalmanson inequalities. This implies some constraints on the behaviour of the characters, with some convexity property and quite stringent parameter evolution along the tree \citep{TF09}.

We are considering here only mathematical criteria to select the parameters that could yield a robust phylogenetic tree. Physical considerations should be avoided as much as possible at this level since our astrophysical a priori conception would bias this choice.

\subsection{Interpretation}

\begin{figure}[ht]
\begin{center}
\includegraphics[height=7.5cm]{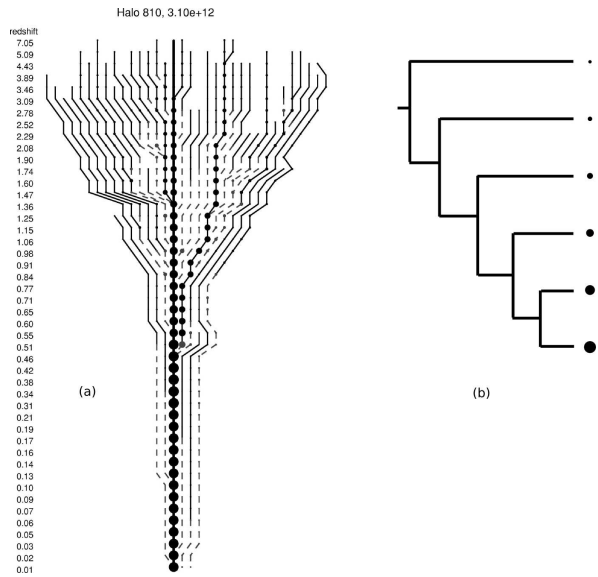}
\caption{\label{fig:DMtrees}Two representation of the diversification of Dark Matter Haloes. (a) is the usual merger tree similar to a genealogy. (b) a cladogram made with one parameter, the size or the mass of the haloes, and represents a phylogeny.} \end{center}
\end{figure}

 Representating astrophysical objects on a tree is quite unusual, except for the evolution of Dark Matter Haloes represented on a graph known as a merger tree \citep[Fig.~\ref{fig:DMtrees}a, from ][]{Stewart2008}. In these cosmological trees, the first seeds of Dark Matter were very small haloes (small both in mass and dimension) that merged under the action of gravity to form bigger and bigger haloes. Hence, from the top of the figure, we obtain downward the genealogy of the bottom big halo.

However, there are many such big haloes, and many haloes with all kinds of sizes. The merger tree representation is unable to show them all, and even to show the evolution of Dark Matter haloes as a whole, as a population. With this kind of graph, how could you depict the evolution of the size of the haloes as a function of time or evolutionary stage of the Universe? This is imposslble. 

In Fig.~\ref{fig:DMtrees}b, the cladogram we would obtain with a sample of Dark Matter haloes using one parameter (mass or dimension) is shown. This is a phylogeny depicting how the species represented by each mass appeared in the course of the evolution of the Universe: most primitive haloes were small, the bigger ones occured progressively by sucessive mergers of smaller ones.

\section{Clustering vs phylogenetic approaches}

\subsection{A simple example}

\begin{figure}[h!]
\begin{center}\includegraphics[width=7 cm]{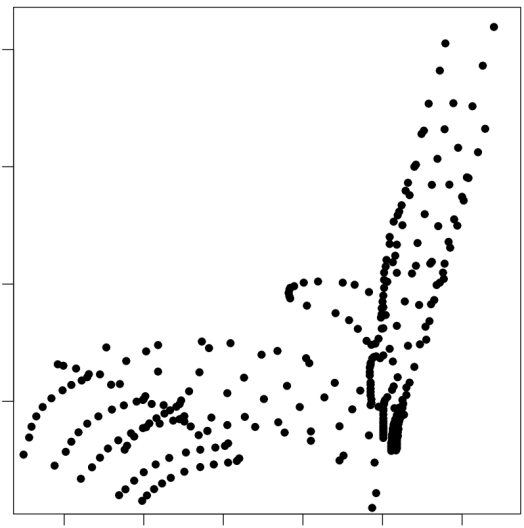}\end{center}
\caption{\label{fig:HR1}A bivariate diagram.}
\end{figure}

How many groups can you identify in the plot of Fig.~\ref{fig:HR1}? 

\begin{figure}[h!]
   \begin{center} \includegraphics[width=7cm]{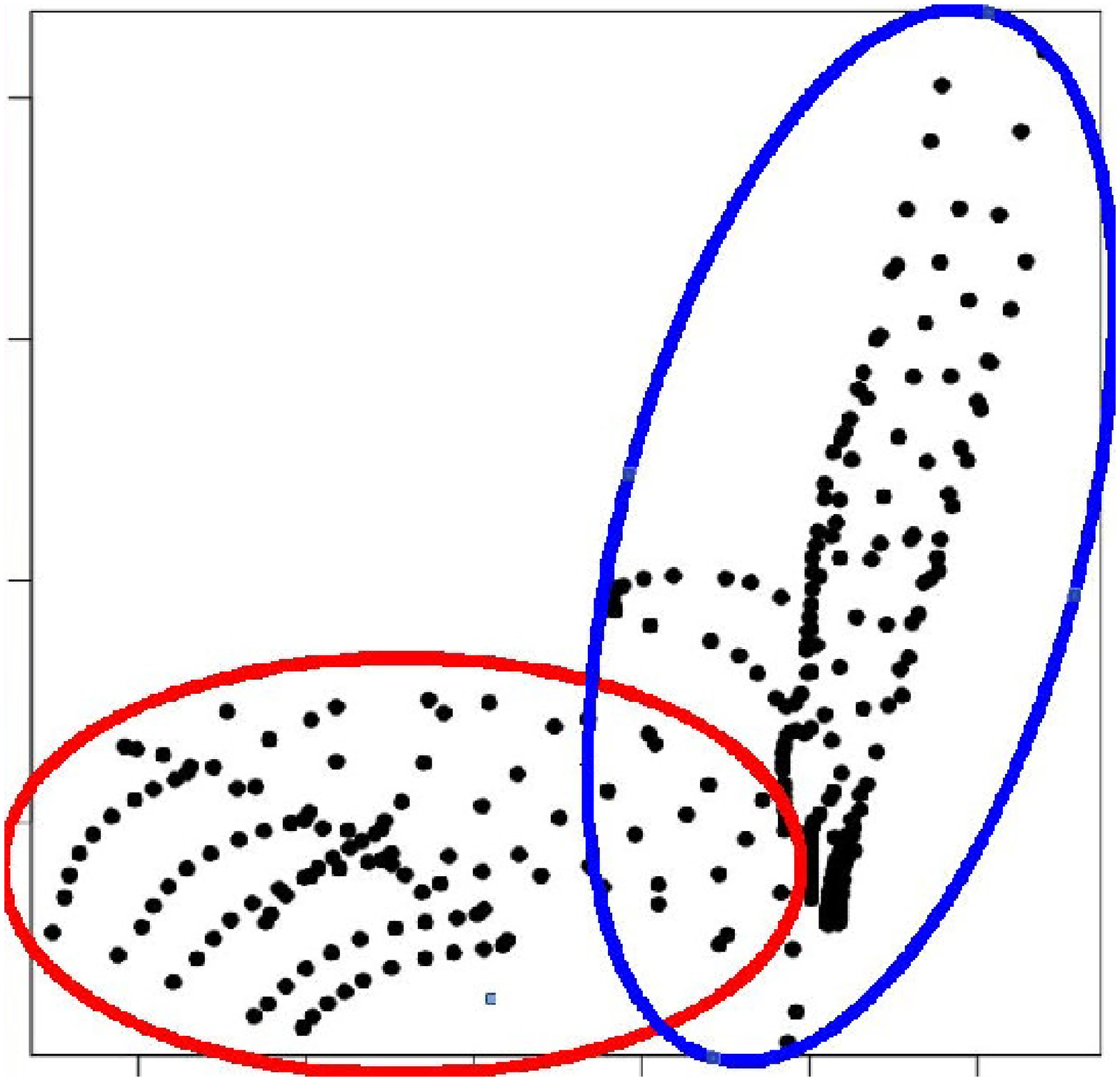}\end{center}
   \caption{\label{fig:HR2}A bivariate diagram where two groups are easily seen.}
\end{figure}

I bet you have said two groups as depicted in Fig.~\ref{fig:HR2}, each one corresponding to different correlation properties.

\begin{figure}[h!]
 \begin{center}\includegraphics[width=8cm]{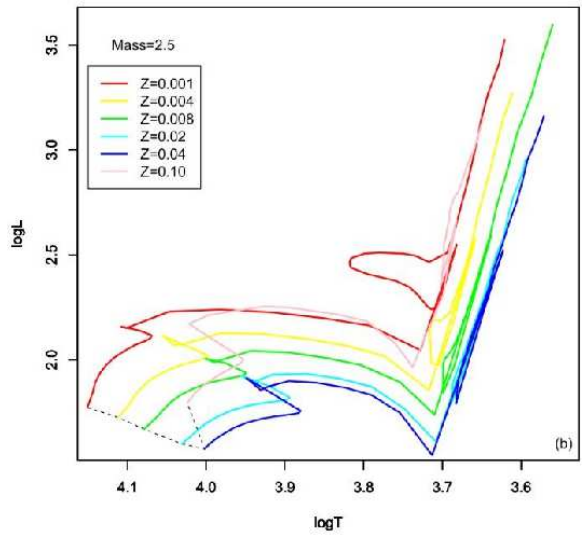}\end{center}
   \caption{\label{fig:HR3}Six stellar evolutionary tracks.}
\end{figure}

The real answer is that there are six populations in this Hertzsprung-Russell (HR)  diagram  showing the luminosity $\log L$ as a function of the surface temperature of the star $\log T_{eff}$ (Fig.~\ref{fig:HR3}). The data come from the Geneva stellar evolutionary models \citep{grid1,grid2,grid3,grid4,grid5} that compute the stellar parameters for stars of masses from M = $0.8$ to $120$ Mo and metallicities Z from $= 0.001$ to $0.1$. In the present sample,  each of the six tracks chosen is represented by 51 time steps. 

The fifteen parameters available are given in Table~\ref{tab:HR} (note that by definition X+Y+Z=1). We will not use the age and the mass of the stars as these parameters are generally not known in practice. Only two ($\log L$ and $\log T_{eff}$) or thirteen (all except age and mass) parameters will be used in the following.

\begin{table}[ht]
 \renewcommand{\arraystretch}{1.2}
\begin{center}
\begin{tabular}{lll}
\hline
 & Parameter & Description \\
 \hline
 & Age     & Age \\
 & Mass    & Actual mass \\
1 & log L  & log(luminosity) \\
2 & log T  & log(effective temperature) \\
3 & Z      & Star metallicity \\
4 & X      & H surface abundance \\
5 & Y      & He surface abundance \\
6 & C12    & $^{12}$C surface abundance \\
7 & C13    & $^{13}$C surface abundance \\
8 & N14    & $^{14}$N surface abundance \\
9 & O16    & $^{16}$O surface abundance \\
10 & O17   & $^{17}$O surface abundance \\
11 & O18   & $^{18}$O surface abundance \\
12 & Ne20  & $^{20}$Ne surface abundance \\
13 & Ne22  & $^{22}$Ne surface abundance \\
 \hline
\end{tabular}
\end{center}
\caption{\label{tab:HR}}
\end{table}

\subsubsection{Partitioning with six groups and phylogenetic analyses}

\begin{figure}[ht]
   \begin{center}
\includegraphics[width=\linewidth]{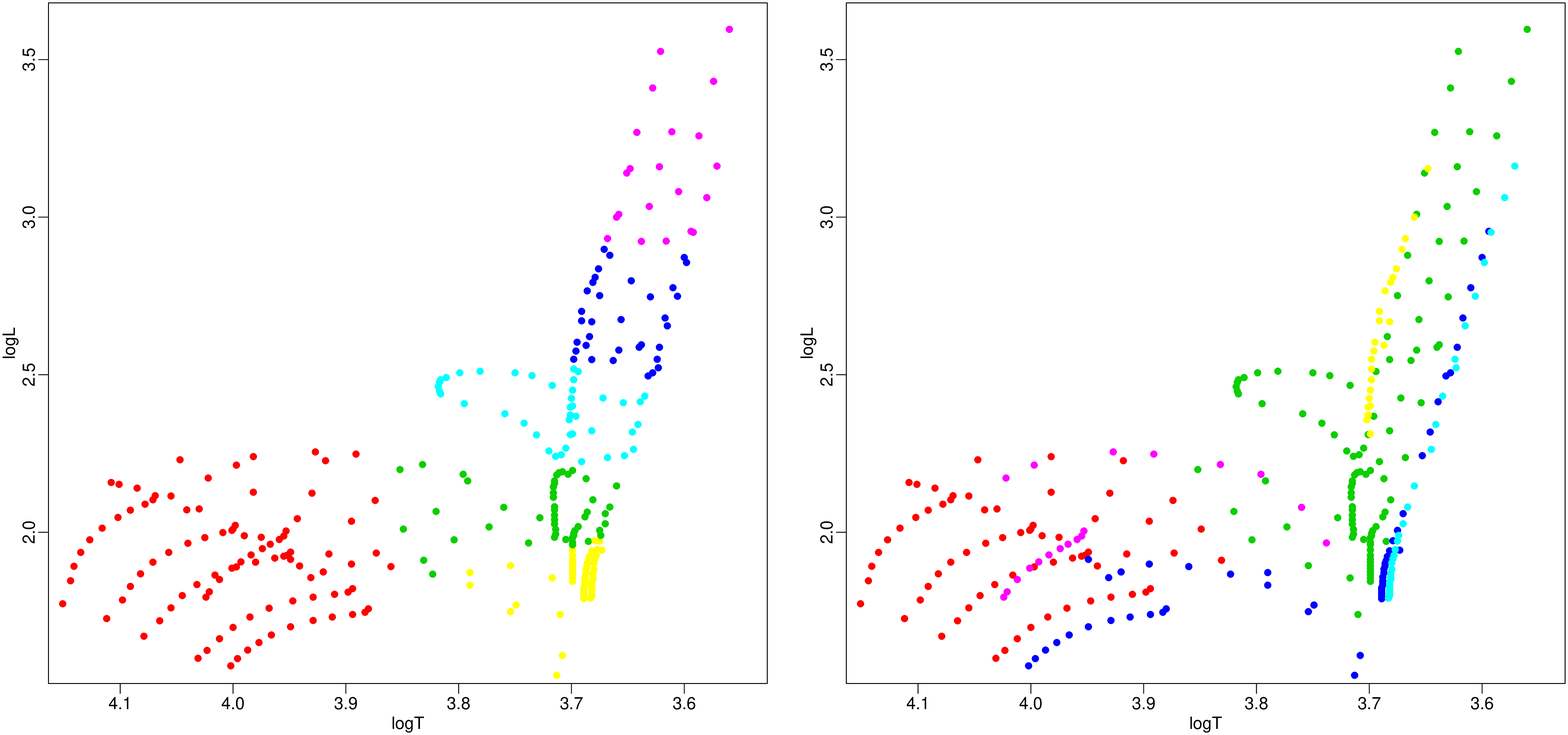}
\end{center}
\caption{\label{fig:HRkmedoids}K-means analysis assuming six groups, with logL and logT only (left) or with the thirteen parameters (right). }
\end{figure}

   Knowing that we should find six groups, it is easy to perform a k-means analysis. The result is shown in Fig.~\ref{fig:HRkmedoids}.

With two parameters, it is clear that the partitioning approach looks for hyperspheres and cannot find elongated or one-dimensional structures. It is sligthly better with thirteen parameters, but still far from satisfactory.

\begin{figure}[ht]
\begin{center}
 \includegraphics[width=0.49\linewidth]{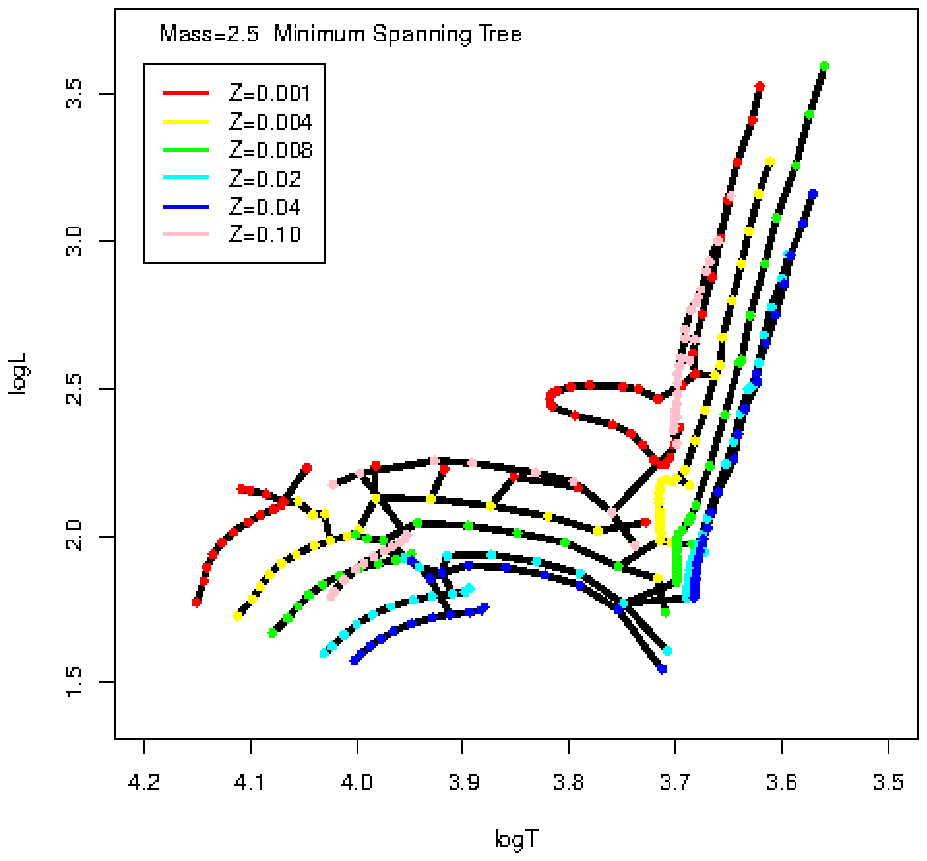}        
 \includegraphics[width=0.49\linewidth]{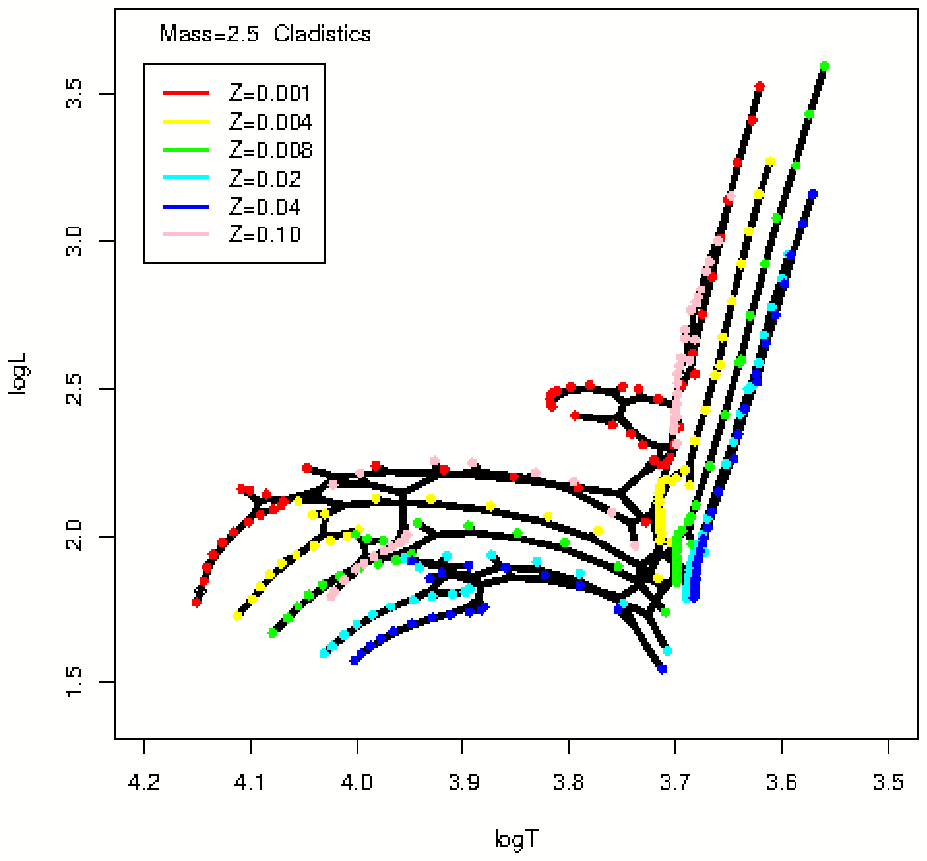}                                                                                     \end{center}
 \caption{\label{fig:HRphylo}Phylogenetic analyses of the stellar sample with thirteen parameters. Left: MST result. Right: MP result. These representations are projections of the trees on the two dimension diagram so that the lines are the branches of the trees.}
      \end{figure}

On the contrary, the phylogenetic methods, MST and MP, perform extremely well as can be seen in Fig.~\ref{fig:HRphylo}. However, this is still not perfect since the tracks are not connected on the lower left at the level of the Main Sequence of the stellar evolution. It is enlightning to investigate this point further as it shows that a clustering analysis always depend on the data and in particular on the available parameters.

\subsubsection{Role of parameter behaviour}

\begin{figure}[ht]
\begin{center}
\includegraphics[width=\linewidth]{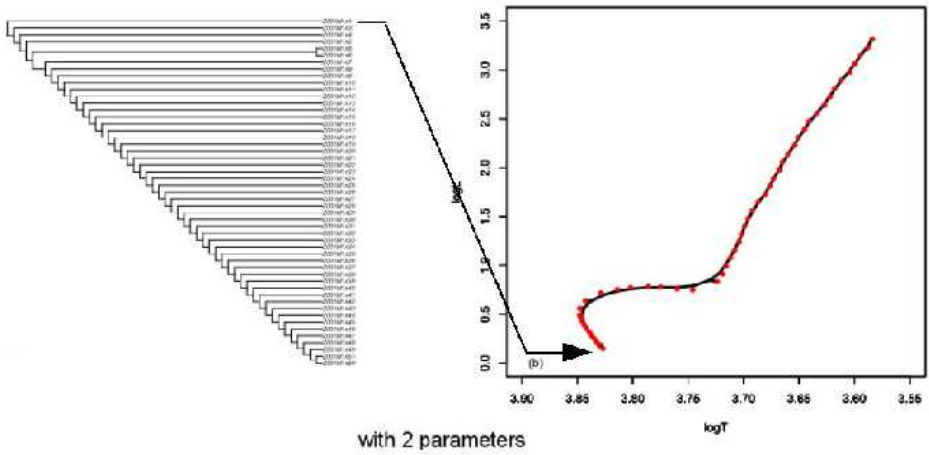}
\end{center}
 \caption{\label{fig:HR1track}MP reconstruction of one stellar track with two parameters (logL and logT). The tree on the left has been rooted with the star the closest to the Main Sequence corresponding to the initial step of the evolution.}
      \end{figure}

Consider one evolutionary track (M=1 Mo, Z=0.001). Only two parameters (logL and logT) are enough here since the reconstruction is easy (Fig.~\ref{fig:HR1track}). 

      The tree is linear, and if rooted correctly, reproduces perfectly the chronology of the stellar evolution. Note that this track is simple, with no loop, there is thus no reversal in any of the two parameters.

\begin{figure}[ht!]
\begin{center}
\includegraphics[width=\linewidth]{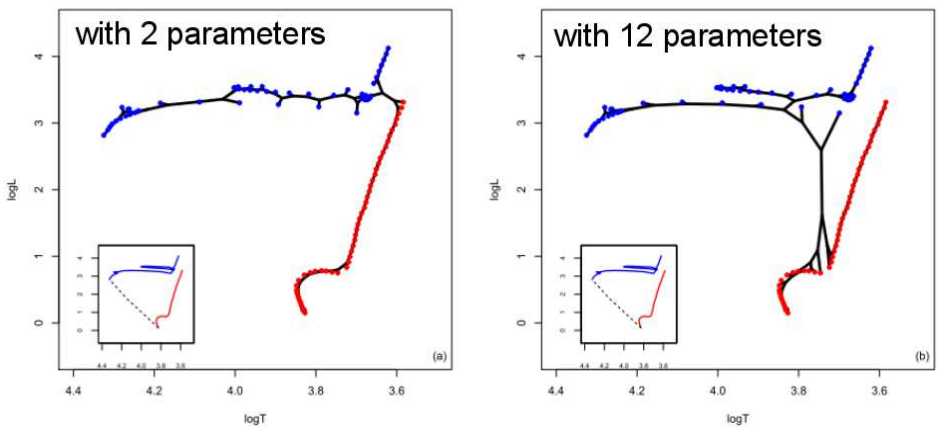}
 \caption{\label{fig:HR2tracks}MP reconstruction of two stellar tracks with two parameters (logL and logT) and twelve (Z is the same for the two tracks).}
 \end{center}
      \end{figure}

Consider now two tracks (M=1 and 5 Mo, Z=0.001) that are shown as they should be in the inlet of Fig.~\ref{fig:HR2tracks}. Note that there are only twelve parameters since the metallicity Z is constant in the two tracks.

There are several problems with the result. The first one is the connection between the two tracks that occurs at the end of them with two parameters, and in the middle with twelve parameters.  This implies that the chronology is not respected at least on of the branches of the tree.  The second problem is the loop seen on the track on the top. This loop is totally missed with two parameters and is only sketched with twelve parameters. 

\begin{figure}[ht]
\includegraphics[width=\linewidth]{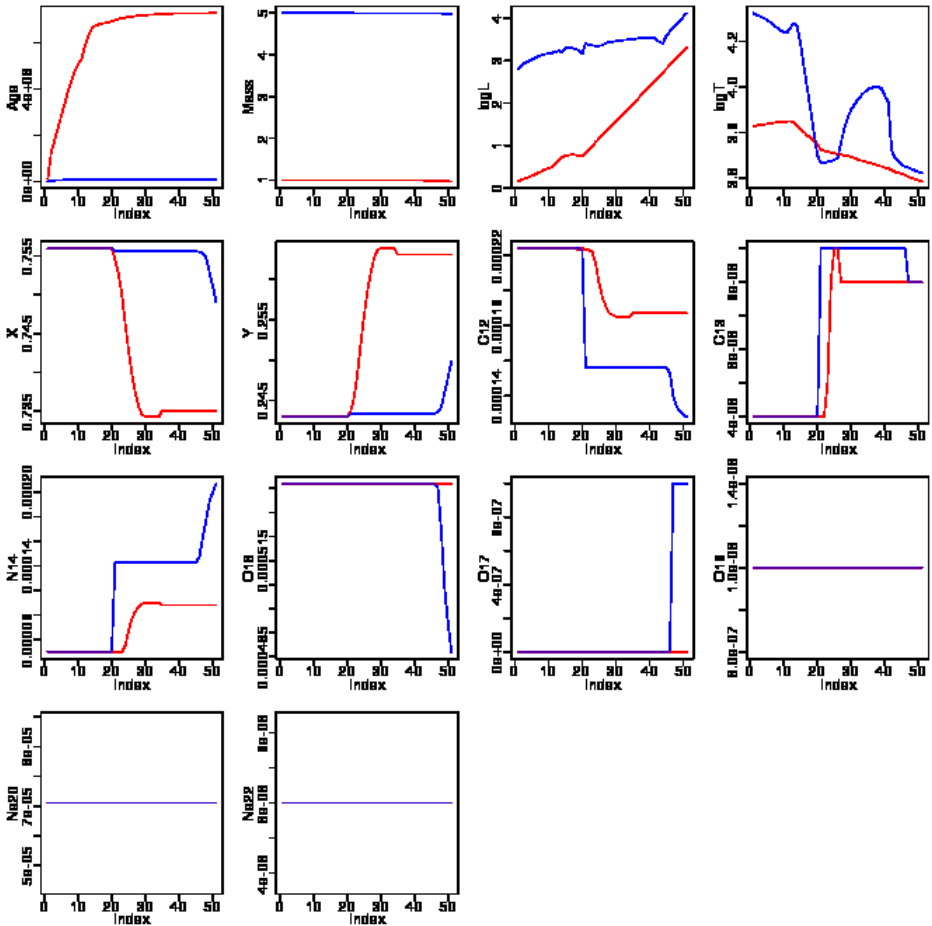}
\caption{\label{fig:parevol}Evolution of the thirteen parameters along the two tracks of Fig.~\ref{fig:HR2tracks}. Note the strong reversal in logT (upper right diagram) and the lack of evolution and often identical values of many parameters at the beginning of the tracks. }
\end{figure}

What should we obtain? Ideally something that looks like the curves in the inlet with a connection at the main sequence (dotted line). Why this is not the case? Let us examine the evolution of the parameters along the track (Fig.~\ref{fig:parevol}).

We can first notice one homoplasy, here a reversal, in logT. Then there are ranges (steps 1 to 20) where the tracks cannot be distinguished on most of the parameters. These behaviours explain why it is not possible to recover the perfect sequences, whatever the technique used.

\section{Astrocladistics in Practice}
\subsection{Workflow}

\begin{figure}[ht]
\begin{center}
   \includegraphics[height=8cm]{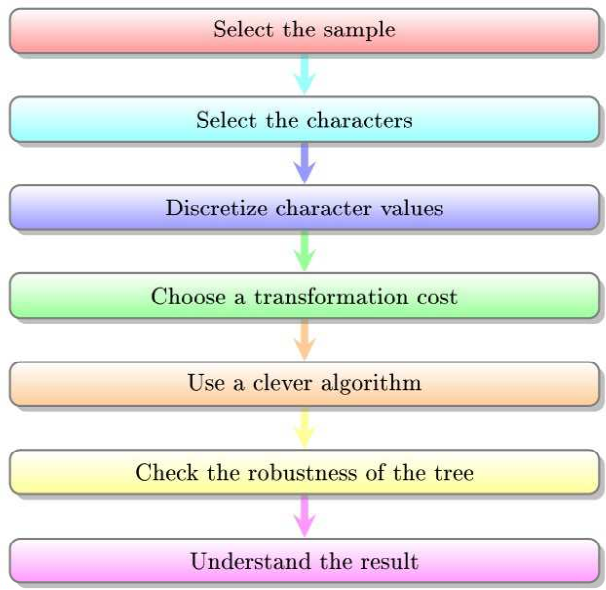}
\end{center}
    \caption{\label{fig:workflow}The workflow of a typical astrocladistic analysis.}
 \end{figure}

We have now seen all the aspects of a complete cladistic analysis. This is summarized in Fig.~\ref{fig:workflow}. It is important to recall that the selection of the sample and of the parameters must be made on statistical grounds, avoiding a priori and subjective choices that will heavily influence the results. For instance, if you choose only structural parameters, you will very certainly obtain a sequence according to what can be called the size.

\subsection{Softwares}

The last point to address is which software could we use to perform the tree search. 

The bioinformaticians have worked for long on sophisticated algorithms to find the phylogenetic classification of living organisms. We can simply use them. All softwares presented below are freely available and most are open source, except PAUP that is in the process of becoming so.

Undoubtly, the most useful software package to undertake a maximum parsimony search on astrophynomical data is \textbf{PAUP*} \citep[Phylogenetic Analysis Using Parsimony\footnote{http://paup.csit.fsu.edu/}, ][]{paup}. There are many useful tools to analyse, manipulate, and modify trees. But an important point I think is the possibility to use up to 32 or 64 bins for the characters. 
     
A very convenient environment to prepare the matrix and analyse the results is the statistical package \texttt{R} which sees a recent and considerable development for phylogenetic studies\footnote{https://www.cran.r-project.org/web/views/Phylogenetics.html}. Unfortunately, it still lacks the ability to perform the MP search itself, even though the package \textit{phangorn} allows this with up to eight bins, which I do not find enough.
    
    With these two softwares, you have all that is necessary to perform a complete cladistic analysis of astrophysical objects.
    There are other tools you may find interesting. 

\textbf{Mesquite}\footnote{http://mesquiteproject.wikispaces.com/} \citep{mesquite} is a rich java environment with very convenient graphical visualisation of trees. There are more and more modules allowing many kinds of computations, and it is linkable with PAUP and \texttt{R} to perform other analyses. 
    
\textbf{MacClade} is as popular and powerful as PAUP but it is only available for MACs and seems not to work anymore on new versions of the OS. They advise to use Mesquite instead. \textbf{Phylip}\footnote{http://evolution.genetics.washington.edu/phylip.html} is also very powerful but devoted mainly to genetic data which have no equivalent in astrophysics. Finally, \textbf{TNT} \citep[Tree analysis using New Technology\footnote{http://www.lillo.org.ar/phylogeny/tnt/}][]{TNT2008} which is more efficient for big data sets (say 500) and also for continuous data that can be binned into 65~000 bins! However there seems to have some assumptions for the changes of character states which may not be adapted to astronomical processes.

\subsection{An application of cladistics in astrophysics: the Globular Clusters of our Galaxy}

\subsubsection{Context}
Globular Clusters are compact and self-gravitating clouds of stars. For each cluster, and in a first approximation, stars are  formed together at the same time in the same physical and chemical conditions. Globular Clusters exist around and in all galaxies, and they are certainly strongly linked to the tranformation processes that build the observed galaxies. Hence finding the different populations of globular clusters for a particular galaxy yields information on its history.

For our own Galaxy, several studies have agreed upon two or three populations. The first natural discriminant parameter is the metallicity ([Fe/H]) that measures the amount of ``heavy''  atomic elements (indeed all except  hydrogen). But it appears that another parameter is needed to explain the diversity of the globular clusters around our Galaxy. This mysterious variable is called the second parameter.

Finding the populations is here a problem of clustering with two parameters, one of them being unknown and not necessarily observable. Such a multivariate problem is usually tackled in astrophysics by defining arbitrary cuts in a scatterplot, i.e. a 2D parameter space. For the particular case of the Glubular Clusters of our Galaxy, one possible second parameter is the distance to the galactic centre. A very good illustration is given in Table~1.6 by \citet{Harris2001}. Subgroups are defined by first splitting the metallicity at -1 and then the distance to galactic centre at 4,8, 12 and 20 kpc. They also can be defined by several splits in the metallicity (-1.85, -1.65, -1.50, -1.32). For each subgroup, the mean of the rotation velocity and its dispersion is computed to conclude on the dynamical differences between these subgroups. It is however hard to believe that hard frontiers in distance from the galactic centre between the globular cluster populations result from a long evolution with many transformation processes of our Galaxy. 

In reality, this is not this distance that matters to explain the different properties of the populations, but the environment where they formed. 

  \subsubsection{The Cladistics Analysis}

If the populations of globular clusters were formed in different environments, they probably formed during different stages of evolution of the environment in our Galaxy at different stages of development. In other words, the environments in which the globular cluster populations formes, were different and are related by ``evolutionary'' relationships due to the changes in metallicities and/or physical conditions. As a consequence, the globular cluster populations are also related by some ``evolutionary'' relationships. 

This problem has been reconsidered with cladistics by \cite{FDC09}. The data were retrieved from the literature on 54 globular clusters of our Galaxy, described by the three parameters selected from previous other studies :
\begin{itemize} 
\item  metallicity ([Fe/H]), 
\item absolute V magnitude (Mv, more or less indicative of the total stellar mass of the cluster),
\item maximum effective temperature (Te) on the horizontal branch.
\end{itemize}

   Each parameter was discretized into 10 equal bins. The number of bins varied (from 3, 5, 8, 15 and 20), and an excellent agreement was found between the results except for 3 and 5 bins. 
The only reasonable evolutionary model here is the Wagner one (ordered states).

\subsubsection{The tree}

\begin{figure}[ht]
   \includegraphics[width=0.35\textwidth]{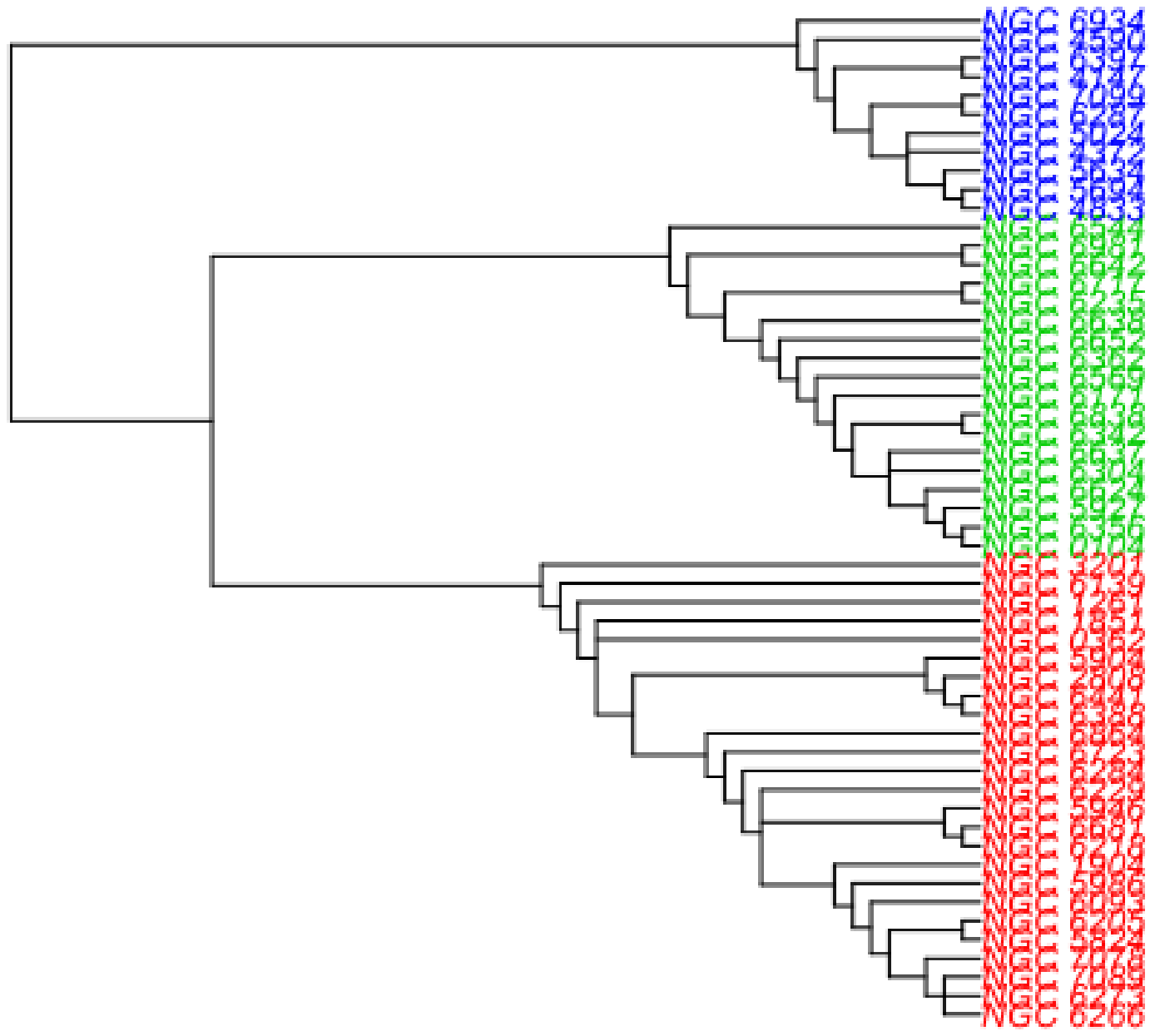}
\includegraphics[width=0.65\textwidth]{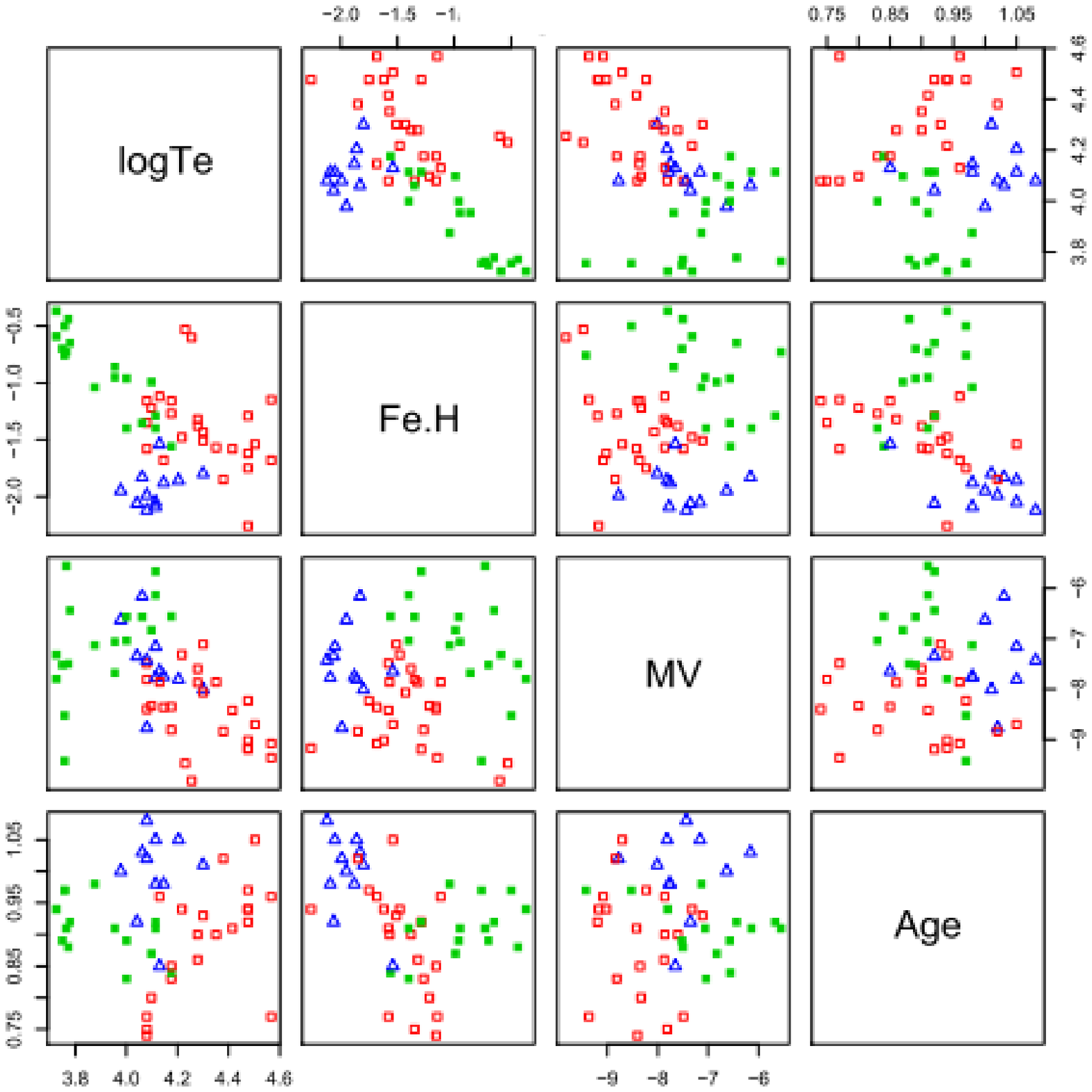}
\caption{\label{fig:AG}Left: the cladogram of the Globular Clusters with the three obvious groups. Right: Scatter plots involving the three parameters used in the analysis plus the age.}
\end{figure}

A very robust tree was found, with three obvious groups shown in Fig.~\ref{fig:AG}. The tree is rooted with the less metallic objects (supposedly the oldest ones since the global metallicity in the Universe increases). The scatterplots in Fig.~\ref{fig:AG} shows the three parameters of the cladistic analysis plus the age that helps to date the groups and the transforming processes of the Galaxy during which the globular clusters formed. It is already easy to derive the different histories for the three groups.

\subsubsection{Interpretation}
   
One can use any parameter available to interpret the result. For instance we find that the three groups have quite different average distances to the galactic centre, as well as different kinematics around our Galaxy. Note that the different distances to the galactic centre, not used in the cladistics analysis, is derived from our analysis, not hypothesized like in the work by \cite{Harris2001} mentioned above.

Using all the information at our hand, we find that the three groups can satisfactorily be interpreted in the frame of our evolutionary hypothesis in the following way (given here for the simple picture of a monolitic assembly of our Galaxy):
\begin{itemize}
   \item {Group 2} (top of the tree, triangle symbols in Fig.~\ref{fig:AG}) formed early during a dissipationless phase of the assembly of our Galaxy, a gentle collapse of a huge cloud of gas,
\item {Group 1} (middle of the tree, filled squares) formed later in a dissipational phase, implying more turbulent and more metallic gas,
\item {Group 3} (bottom of the tree, open squares) formed in the disk of our Galaxy and rapidly in an intermediate epoch.
\end{itemize}

\subsection{To go further}

Astrocladistics is the implementation of phylogenetic techniques in astrophysics. Three fundamental papers \citep{FCD06,jc1,jc2} explain in detail why and how cladistics can be used in the case of galaxies, using samples taken from cosmological numerical simulations or closeby Dwarf galaxies. 

You can find other applications of cladistics on galaxies \citep{Fraix2010,Fraix2012}, globular clusters \citep{FD15} and Gamma Ray Bursts \citep{Cardone2013}.

\section{Generalization: Networks}

 \subsection{Split networks}

 Outer planar networks are generalizations of trees \citep{Huson2006} since they can represent simultaneously alternative trees with reticulations (generated for instance by hybridization). The link with phylogenetic trees is given by the definition of splits \citep[see e.g.][]{Fraix-Burnet2015}. 
 
\begin{figure}[ht]
\begin{center}
 \includegraphics[width=4 cm]{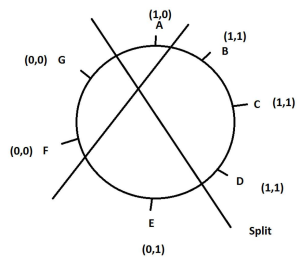}                                                \end{center}
 \caption{\label{fig:split}A circular order for objects A to G, with their pairs of binary states. The two lines show two different splits: (0,) vs (1,) and (,0) vs (,1).}
\end{figure}

 A split creates a partition of objects into two disjoint sets. Objects sharing a common property, as defined by splits, are consecutive in a circular order. This can be more easily visualized for binary characters (Fig.~\ref{fig:split}). Note that multistate characters can be transformed into binary characters. 

\begin{figure}[ht]
 \begin{center}
 \includegraphics[width=\textwidth]{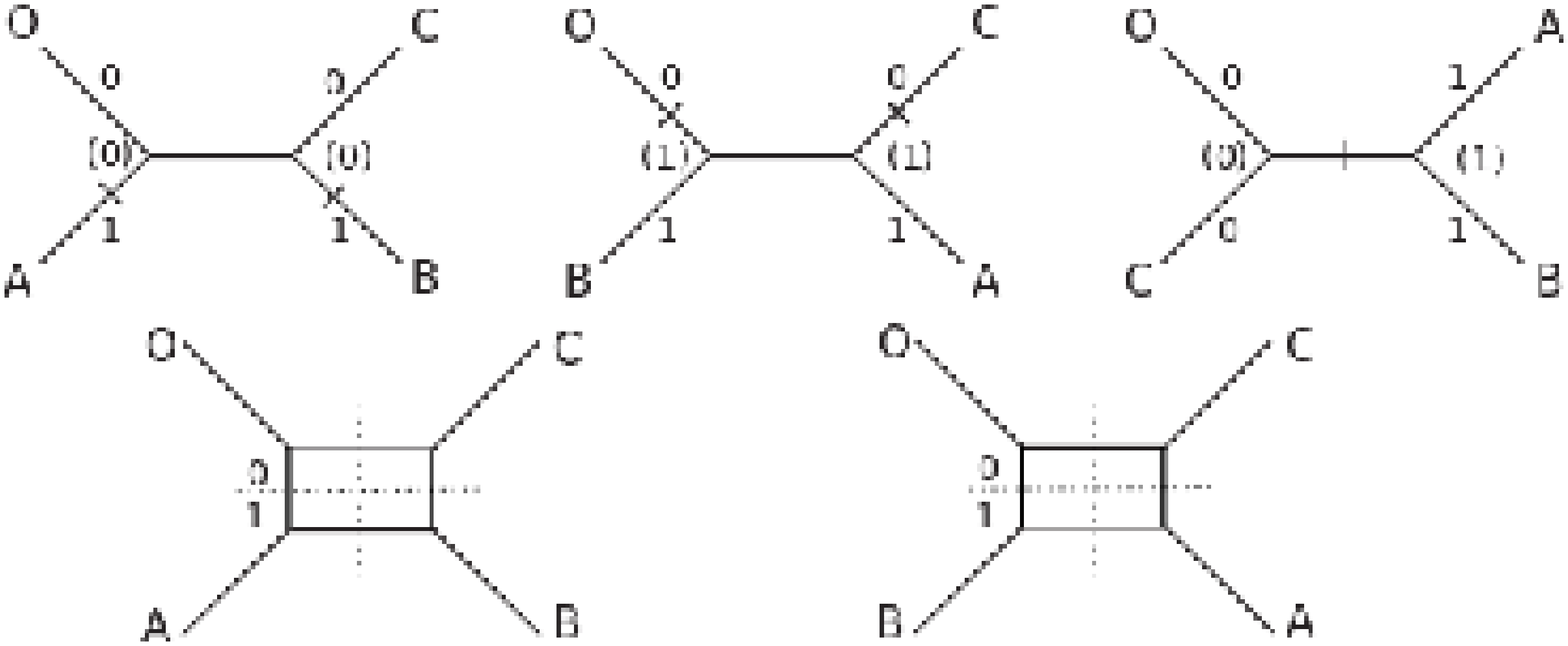}                                                                    \end{center}
 \caption{\label{fig:multtree}Top: the three possible arrangements of four objects described by one binary character. Bottom: the two split networks combining two pairs of the three trees. The dotted lines indicate the split with the corresponding values of the character.}
\end{figure}

In other words, splits in an outer planar network furnish neighboring relationships between objects. Figure~\ref{fig:multtree} explains how several trees are represented by a network and how splits separate the different trees. With four objects, there are three possible tree arrangements. Adding transversal branches allows to combine two of these trees, that is to merge two conflicting trees into a single diagram. A 3-dimensional representation could combine the three trees on a single scheme.

\begin{figure}[ht]
\begin{center}
 \includegraphics[width=0.9\textwidth]{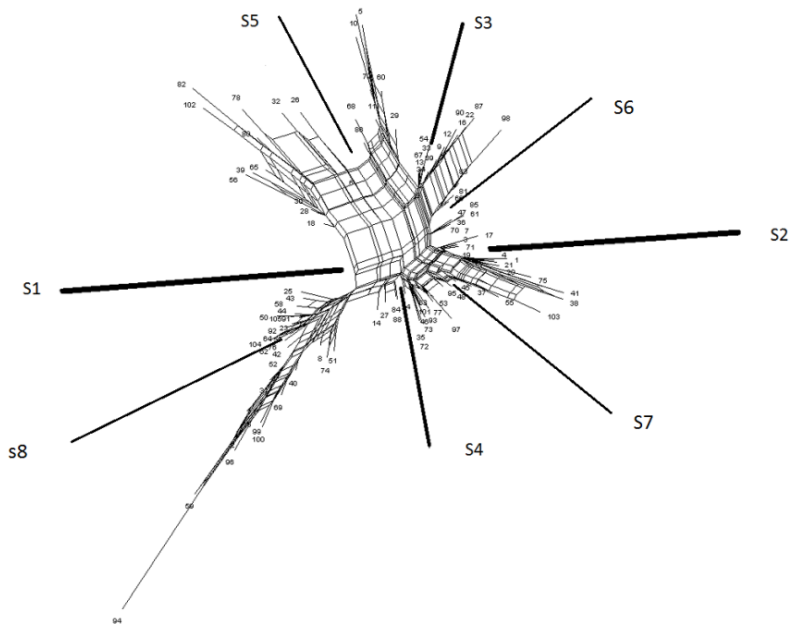}                                                            \end{center}
 \caption{\label{fig:network}An example of an outer planar network showing the eight splits of the eight parameters s1 ... s8.}
\end{figure}

Networks can be very complicated (Fig.~\ref{fig:network}) and become rapidly quite difficult to read, and to interpret. However, this is probably the closest scheme to the true diversification scenario of living organisms, especially bacteria. It is not impossible that we also should use this kind of representation for galaxy diversification. You can find an example of its potential use in astrophysics in \citet{TF09}.

\section{Conclusion}

 In a multivariate world, in the era of big data, we need adapted methodologies. 
 
 With the evolution of our knowledge and our observational/experimental technologies, we need to modify our tools for data analysis. In particular, our classifications must evolve accordingly.

 Evolution, which is the continuous transformation of objects, is inherent to any astrophysical object and population.
 
 Hence, the use of phylogenetic approaches is justified and must be pursued. This is the goal of astrocladistics\footnote{\url{http://astrocladistics.org/}} that started fifteen years ago.

\bigskip

Astronomers, as physicists, put physics above everything else. But before doing science with your data, don't you process them to correct for instrumental distortions? Don't you transform them to get calibrated quantitative information? Don't you use signal processing techniques to diminish the noise, extract some hidden signal or translate it into something tractable? Doing advanced statistical analyses proceeds to the same kind of data management that aims at finding the information that can be used for the physical interpretation of the data.

 In any statistical clustering study, be it phylogenetic or not, you need to be rigorous: 
\begin{itemize}
   \item Know your data (\emph{statistics})
   \item Select your parameters objectively (\emph{statistics})
   \item Compare several methods (\emph{statistics})
   \item Characterize your groups (\emph{statistics})
   \item And then, only then, interpret (\emph{astrophysics})
\end{itemize}

Yes astrophysics comes last, to interpret a statistical robust result, and to conclude whether the statistical analysis brings something useful. I like to compare the statistical tools as telescopes that help us explore an invisible world.

\Appendix

\section[Exercices]{Exercices\footnote{The corresponding \texttt{R} codes and the required data files can be found at \url{http://stat4astro2015.sciencesconf.org/}}}

\subsection{Presentation of the session}

In this session you will compute and play with phylogenetic trees obtained by Maximum Parsimony (MP or cladistics), Minimum Spanning Tree (MST) and Neighbor Joining Tree Estimation (NJ).

The computation of Maximum Parsimony in \texttt{R} is not yet as powerful as it is in other dedicated softwares. However the package phangorn, mainly intended for genetic data, will be sufficient for this session.

\subsection{Packages to install}

You need to install these packages, either from the command line or from RStudio. If you know your repository, add the option: repos=``http://...'' .
\begin{lstlisting}[language=R]
install.packages("ape")      # the basis for most 
                             #      phylogenetic work
install.packages("phytools") # a new set of tools
install.packages("phangorn") # for computing Maximum 
                             #   Parsimony and other stuff
install.packages("igraph")   # the network analysis package
\end{lstlisting} 

Then load the three first libraries:
\begin{lstlisting}[language=R]
library(ape)
library(phytools)
library(phangorn)
\end{lstlisting}

\subsection{MP analysis}

\subsubsection{A simple example}

The file Data/phylo\_data.txt contains an artificial data set of 10 galaxies with 7 binary properties: ``0'' means either low or absent, ``1'' means high or present.

Load the matrix and transform it to the class phyDat for use in phangorn:

\begin{lstlisting}[language=R]
X <- read.table("./Data/phylo_data.txt", header = TRUE)
X
X <- as.phyDat(X, type = "USER", levels = c(0, 1))
X
\end{lstlisting}

First check all the possible trees with 4 and 5 objects:
\begin{lstlisting}[language=R]
trees <- allTrees(4) ; trees 
 op <- par(mfrow = c(1,3))  
 for(i in 1:length(trees)) plot(trees[[i]],type="unrooted",
                                         cex=1.5) ; par(op)
trees <- allTrees(5) ; trees 
 op <- par(mfrow = c(3,5)) 
 for(i in 1:length(trees)) plot(trees[[i]],type="unrooted",
                                         cex=1.5) ; par(op)
\end{lstlisting}

How many trees are there for 6 objects? For 7?

We use the ratchet technique to compute the MP tree without exploring all possible arrangements. This is a very efficient method that avoids as much as possible to be trapped in a local minimum. Do not worry about the parameters of the command, it is not important for us here.
\begin{lstlisting}[language=R]
tree <- pratchet(X,method="sankoff",all=TRUE,maxit=100000,
                                                    k=100)
 op <- par(mfrow=c(2,2))
plotTree(tree)  ; par(ask=F)
 par(op)
\end{lstlisting}

Why are there several trees?

There are two ways to summarize the multiple trees. The first one is to build a consensus tree. It can be a strict consensus (nodes occuring in 100\% of the trees) or a majority rule (always higher than 50\%, 70\% is a good compromise).
\begin{lstlisting}[language=R]
tcons <- consensus(tree,p=0.7)
plotTree(tcons)
tcons <- root(tcons, outgroup = "GAL1", resolve.root = TRUE)
plotTree(tcons)
\end{lstlisting}

Try to root with another galaxy.

The second way to summarize the multiple trees is to compute a split network. 
\begin{lstlisting}[language=R]
tnet <- consensusNet(tree,prob=0.3)
plot(tnet,"2D")
plot(tnet, "2D",show.edge.label=TRUE)
plot(tnet) # in 3D you need the package rgl
\end{lstlisting}
To better understand the split network, compare with the consensus.
\begin{lstlisting}[language=R]
op <- par(mfrow=c(1,2))
plotTree(tcons)
plot(tnet,"2D")
par(op)
\end{lstlisting}

\subsubsection{Local dwarf galaxies}

Load the data in the file dwarfs.txt and select only 14 galaxies for which we know a robust tree exists.

\begin{lstlisting}[language=R]
dwarfs <- read.delim("Data/dwarfs.txt",na.strings="?")
dwarfs14 <- dwarfs[c(10,7,35,33,15,9,11,16,30,1,25,5,4,24),]
str(dwarfs14)
\end{lstlisting}
To code this matrix, I provide you with a script that you have to source. We choose 8 bins since it is the maximum allowed in phangorn.
\begin{lstlisting}[language=R]
source("Data/CodeMatrix.R")
dwarfs14cod <- CodeMatrix(dwarfs14,bins=8)
Y <- as.phyDat(data.frame(t(dwarfs14cod)), type = "USER", 
                        levels = c(0, 1,2,3,4,5,6,7,"?"))
\end{lstlisting}

Now perform the Maximum Parsimony Analysis as previously. The command to root the tree looks odd (write and read), this is to ensure that the order of the names corresponds to the tree order. This is not absolutely necessary but avoids some confusion in some plotting.  The ladderize function organizes the tree so that it is easier to compare several trees.
\begin{lstlisting}[language=R]
treeMP <- pratchet(Y,method="sankoff",all=TRUE,
                          maxit=100000,k=100) ; treeMP
plotTree(treeMP) ; par(ask=F)
treeMP <- read.tree(text=write.tree(ladderize(root(treeMP,
                                      outgroup="SagDIG"))))
plotTree(treeMP) ; par(ask=F)

\end{lstlisting}

Generally, distance-based approaches do not accept NA values. In order to compare to MST and NJ, we now work with a reduced matrix where all parameters having NA are removed. How many parameters are left?
\begin{lstlisting}[language=R]
dwarfs14red <- dwarfs14[,-c(unique(which(is.na(dwarfs14),
                                          arr.ind=T)[,2]))]
dwarfs14redcod <- CodeMatrix(dwarfs14red,bins=8)
Z <- as.phyDat(data.frame(t(dwarfs14redcod)), type = "USER",
                          levels = c(0, 1,2,3,4,5,6,7,"?"))
treeMPred <- pratchet(Z,method="sankoff",all=TRUE,
                                 maxit=100000,k=100)
treeMPred <- read.tree(text=write.tree(ladderize(root(
                             treeMPred,outgroup="SagDIG"))))
plot(treeMPred)

\end{lstlisting}

Plot side by side the two trees treeMP and treeMPred. Do you see a difference? How can you explain it?

\subsection{Minimum Spanning Tree}

We first need to compute the pairwise distance matrix, and then compute the MST. We can use the mst command of the ape package.

\begin{lstlisting}[language=R]
dis <- dist(scale(dwarfs14red),method="manhattan")

# ape
treemst <- mst(dis)
plot(treemst)
plot(treemst,graph="nsca")
PC <- prcomp(scale(dis))   # PCA analysis
plot.mst(treemst, x1 = PC$x[, 1], x2 = PC$x[, 2])
# in the V vs M diagram:
plot.mst(treemst,x1=dwarfs14red[,3],x2=dwarfs14red[,4])   
\end{lstlisting}

The igraph package is more powerful for graphical representation of networks.
We import the distance matrix into igraph and compute the MST. We can get informations on the list of vertices (nodes), edges (branches), their weigths (distances) and the degrees of the vertices.

\begin{lstlisting}[language=R] 
library(igraph)
net <- graph.adjacency(as.matrix(dis),mode="undirected",
                                    weighted=TRUE,diag=F)

g <- minimum.spanning.tree(net) # or mst(net)
V(g) # gives the list of vertices
E(g) # gives the list of edges
E(g)$weight
degree(g) # gives the number of edges per vertex
\end{lstlisting}

There are many possibilities to plot the tree. In particular there are several ``layout'' that optimize the visibility of all vertices and edges on the graph.
\begin{lstlisting}[language=R]
plot(g, layout=layout_with_fr, vertex.size=4,vertex.
                       label.dist=0.5, vertex.color="red",
                       edge.width=E(g)$weight,
                       edge.label=round(E(g)$weight,2))
# try other layouts : 
# layout_as_tree, layout.auto, layout_with_kk
plot(g, layout=layout.mds(g,dist=as.matrix(dis)), 
                  vertex.size=4,vertex.label.dist=0.5, 
                  vertex.color="red",edge.width=E(g)$weight,
                  edge.label=round(E(g)$weight,2))
\end{lstlisting}

We can also separate the MST by cutting the edges whose length is larger than say 5.
\begin{lstlisting}[language=R]
gp <- g-E(g)[which(E(g)$weight >=5)]
plot(gp, layout=layout.mds(gp,dist=as.matrix(dis)), 
         vertex.size=4,vertex.label.dist=0.5, 
         vertex.color="red",edge.width=E(gp)$weight,
         edge.label=round(E(gp)$weight,2))
\end{lstlisting}

We have finished with igraph, it is good to unload it:
\begin{lstlisting}[language=R]
 detach("package:igraph", unload=TRUE)
\end{lstlisting}

\subsection{Neighbor Joining Tree Estimation}

We use the same pairwise distance matrix as for the MST computation.
\begin{lstlisting}[language=R]
treenj <- nj(dis)
treenj <- read.tree(text=write.tree(ladderize(root(treenj,
                                     outgroup="SagDIG"))))
plot(treenj,use.edge.length=FALSE)
plot(treenj,type="unrooted")
\end{lstlisting}

\subsection{Comparison of the three results}

Analyse the differences between the trees of dwarf galaxies obtained with the three techniques.
\begin{lstlisting}[language=R]
 op <- par(mfrow=c(1,3))
plot(treeMPred,main="MP",cex=1.3)
plot(g, layout=layout_as_tree, vertex.size=4,
                       vertex.label.dist=0.5,main="MST",
                       vertex.label.cex=1.5)
plot(treenj,use.edge.length=FALSE,main="NJ",cex=1.3)
 par(op)
\end{lstlisting}

\subsection{Interpretation of the tree}

There are many ways to project parameter values on the tree, or the tree on diagrams. Here is only two examples using the phytools package. You can now try to identify and characterize some groups from the tree. Try to plot the nine parameters using a loop and the function below for the trees treeMP, treeMPred and treenj.

\begin{lstlisting}[language=R]
interpret <- function(param,tree) {
  layout(matrix(c(1,2),1,2),c(0.7,0.3))
  plot(tree,mar=c(4.1,1.1,1.1,0),use.edge.length=FALSE)
  par(mar=c(4.1,0,1.1,1.1))
  barplot(dwarfs14red[tree$tip.label,param],
            horiz=TRUE,width=1,space=0,
            ylim=c(1,length(tree$tip.label))-0.5,names="")
  title(colnames(dwarfs14red)[param])
}
\end{lstlisting}

There is also a nice way to visualize the evolution of the parameters along the tree.

\begin{lstlisting}[language=R]
evol <- function(param,tree) {
  v <- dwarfs14red[[param]]  
  names(v) <- rownames(dwarfs14red)
  tree$edge.length <- rep(1,length(tree$edge[,1]))
  contMap(tree,v)
  title(colnames(dwarfs14red)[param])
}\end{lstlisting}

You can redo the exercises with the full data set in dwarfs.txt (36 galaxies).


\begin{thebibliography}{31}
\expandafter\ifx\csname natexlab\endcsname\relax\def\natexlab#1{#1}\fi

\bibitem[{Adanson(1763)}]{adanson}
Adanson M. 1763. \emph{Famille Des Plantes}, chez Vincent, impr.-libraire de Mgr le
  Comte de Provence (Paris), num. BNF de l'\'ed. de Paris : INALF, 1961
  (Frantext~; R263R)

\bibitem[{{Barrow} {et~al.}(1985){Barrow}, {Bhavsar}, \& {Sonoda}}]{Barrow1985}
{Barrow} J.~D., {Bhavsar} S.~P. \& {Sonoda} D.~H. 1985. {Minimal spanning trees,
  filaments and galaxy clustering}, \mnras, 216, 17
  
\bibitem[{Cardone \& Fraix-Burnet(2013)}]{Cardone2013}
Cardone V. F. \& Fraix-Burnet D. 2013. Hints for families of GRBs
  improving the hubble diagram, \mnras, 434, 1930

\bibitem[{{Charbonnel} {et~al.}(1993){Charbonnel}, {Meynet}, {Maeder},
  {Schaller}, \& {Schaerer}}]{grid2}
{Charbonnel} C., {Meynet} G., {Maeder} A., {Schaller} G. \& {Schaerer} D. 1993.
  Grids of stellar models - part three - from 0.8 to 120-solar-masses at
  z=0.004, \aaps, 101, 415

\bibitem[{Feigelson \& Babu(2012)}]{FeigelsonBabu2012}
Feigelson E. \& Babu G. 2012. \emph{Modern Statistical Methods for Astronomy: With \texttt{R}
  Applications}. Cambridge University Press

\bibitem[{Felsenstein(1984)}]{Felsenstein1984}
Felsenstein J. 1984. \emph{Cladistics: Perspectives on the reconstruction of
  evolutionary history}, Duncan T., Stuessy T., eds., Columbia University Press,
  New York, pp. 169--191

\bibitem[{{Fraix-Burnet} {et~al.}(2012){Fraix-Burnet}, {Chattopadhyay},
  {Chattopadhyay}, {Davoust}, \& {Thuillard}}]{Fraix2012}
{Fraix-Burnet} D., {Chattopadhyay} T., {Chattopadhyay} A.~K., {Davoust} E. \&
  {Thuillard} M. 2012. A six-parameter space to describe galaxy
  diversification, \aap, 545, A80

\bibitem[{{Fraix-Burnet} {et~al.}(2006{\natexlab{a}}){Fraix-Burnet}, {C}holer,
  \& {D}ouzery}]{FCD06}
{Fraix-Burnet} D., {C}holer P. \& {D}ouzery E. 2006{\natexlab{a}}. {T}owards a
  {P}hylogenetic {A}nalysis of {G}alaxy {E}volution : a {C}ase {S}tudy with the
  {D}warf {G}alaxies of the {L}ocal {G}roup, \aap, 455, 845

\bibitem[{{Fraix-Burnet} {et~al.}(2006{\natexlab{b}}){Fraix-Burnet}, {C}holer,
  {D}ouzery, \& {V}erhamme}]{jc1}
{Fraix-Burnet} D., {C}holer P., {D}ouzery E. \& {V}erhamme A.
  2006{\natexlab{b}}. {A}strocladistics: a phylogenetic analysis of galaxy
  evolution {I}. {C}haracter evolutions and galaxy histories, \emph{{J}ournal of {C}lassification}, 23, 31

\bibitem[{Fraix-Burnet \& Davoust(2015)}]{FD15}
Fraix-Burnet D. \& Davoust E. 2015. Stellar populations in $\omega$ centauri: a
  multivariate analysis, \mnras, 450, 3431

\bibitem[{{Fraix-Burnet} {et~al.}(2009){Fraix-Burnet}, {D}avoust, \&
  {C}harbonnel}]{FDC09}
{Fraix-Burnet} D., {D}avoust E. \& {C}harbonnel C. 2009. {T}he environment of
  formation as a second parameter for globular cluster classification, \mnras, 398, 1706

\bibitem[{{Fraix-Burnet} {et~al.}(2006{\natexlab{c}}){Fraix-Burnet}, {D}ouzery,
  {C}holer, \& {V}erhamme}]{jc2}
{Fraix-Burnet} D., {D}ouzery E., {C}holer P. \& {V}erhamme A.
  2006{\natexlab{c}}. {A}strocladistics: a phylogenetic analysis of galaxy
  evolution {II}. {F}ormation and diversification of galaxies, \emph{{J}ournal of {C}lassification}, 23, 57

\bibitem[{{Fraix-Burnet} {et~al.}(2010){Fraix-Burnet}, {Dugu{\'e}},
  {Chattopadhyay}, {Chattopadhyay}, \& {Davoust}}]{Fraix2010}
{Fraix-Burnet} D., {Dugu{\'e}} M., {Chattopadhyay} T., {Chattopadhyay} A.~K. \&
  {Davoust} E. 2010. Structures in the fundamental plane of early-type
  galaxies, \mnras, 407, 2207

\bibitem[{Fraix-Burnet {et~al.}(2015)Fraix-Burnet, Thuillard, \&
  Chattopadhyay}]{Fraix-Burnet2015}
Fraix-Burnet D., Thuillard M. \& Chattopadhyay A.~K. 2015.  Multivariate
  approaches to classification in extragalactic astronomy, \emph{Frontiers in
  Astronomy and Space Sciences}, 2, 3

\bibitem[{Gascuel \& Steel(2006)}]{NJ2006}
Gascuel O. \& Steel M. 2006. Neighbor-joining revealed, \emph{Molecular Biology and Evolution}, 23, 1997

\bibitem[{Goloboff {et~al.}(2008)Goloboff, Farris, \& Nixon}]{TNT2008}
Goloboff P.~A., Farris J.~S. \& Nixon K.~C. 2008. {T}{N}{T}, a free program for
  phylogenetic analysis,  \emph{Cladistics}, 24.5, 774

\bibitem[{Gower \& Ross(1969)}]{Gower1969}
Gower J.~C. \& Ross G. J.~S. 1969. Minimum spanning trees and single linkage
  cluster analysis, \emph{Journal of the Royal Statistical Society.
  Series C (Applied Statistics)}, 18, pp. 54

\bibitem[{{Harris}(2001)}]{Harris2001}
{Harris} W.~E. 2001. Globular Cluster Systems, Saas-Fee Advanced Course, Vol.~28, \emph{Star Clusters},
  {Labhardt} L., {Binggeli} B., eds., Springer-Verlag Berlin Heidelberg, p.
  223

\bibitem[{Hennig(1965)}]{hennig1965}
Hennig W. 1965, Phylogenetic systematics, \emph{Annual Review of Entomology}, 10, 97

\bibitem[{Huson \& Bryant(2006)}]{Huson2006}
Huson D.~H. \& Bryant D. 2006, Application of phylogenetic networks in
  evolutionary studies, \emph{Molecular Biology and Evolution}, 23, 254

\bibitem[{{Maddison} \& {Maddison}(2004)}]{mesquite}
{Maddison} W.~P. \& {Maddison} D.~R. 2004, Mesquite: a modular system for
  evolutionary analysis (http://mesquiteproject.org)

\bibitem[{{Mowlavi} {et~al.}(1998){Mowlavi}, {Schaerer}, {Meynet},
  {Bernasconi}, {Charbonnel}, \& {Maeder}}]{grid5}
{Mowlavi} N., {Schaerer} D., {Meynet} G., {Bernasconi} P.~A., {Charbonnel} C. \&
  {Maeder} A. 1998, Grids of stellar models. vii. from 0.8 to 60
  m\_$\backslash$odot at z = 0.10, \aaps, 128, 471

\bibitem[{Saitou \& Nei(1987)}]{NJ1987}
Saitou N. \& Nei M. 1987, The neighbor-joining method: a new method for
  reconstructing phylogenetic trees, \emph{Molecular Biology and Evolution}, 4, 406

\bibitem[{{Schaerer} {et~al.}(1993{\natexlab{a}}){Schaerer}, {Charbonnel},
  {Meynet}, {Maeder}, \& {Schaller}}]{grid3}
{Schaerer} D., {Charbonnel} C., {Meynet} G., {Maeder} A. \& {Schaller} G.
  1993{\natexlab{a}}, Grids of stellar models - part four - from 0.8-solar-mass
  to 120-solar-masses at z=0.040, \aaps, 102, 339

\bibitem[{{Schaerer} {et~al.}(1993{\natexlab{b}}){Schaerer}, {Meynet},
  {Maeder}, \& {Schaller}}]{grid4}
{Schaerer} D., {Meynet} G., {Maeder} A. \& {Schaller} G. 1993{\natexlab{b}},
  Grids of stellar models. ii - from 0.8 to 120 solar masses at z = 0.008,\aaps, 98, 523

\bibitem[{{Schaller} {et~al.}(1992){Schaller}, {Schaerer}, {Meynet}, \&
  {Maeder}}]{grid1}
{Schaller} G., {Schaerer} D., {Meynet} G. \& {Maeder} A. 1992, New grids of
  stellar models from 0.8 to 120 solar masses at z = 0.020 and z = 0.001, \aaps, 96, 269

\bibitem[{{Stewart} {et~al.}(2008){Stewart}, {Bullock}, {Wechsler}, {Maller},
  \& {Zentner}}]{Stewart2008}
{Stewart} K.~R., {Bullock} J.~S., {Wechsler} R.~H., {Maller} A.~H. \& {Zentner}
  A.~R. 2008, Merger histories of galaxy halos and implications for disk
  survival, \apj, 683, 597

\bibitem[{Swofford(2003)}]{paup}
Swofford D.~L. 2003, Paup*: Phylogenetic analysis using parsimony (*and other
  methods), nauer Associates, Sunderland, Massachusetts, {http://paup.csit.fsu.edu/}

\bibitem[{{T}huillard \& {Fraix-Burnet}(2009)}]{TF09}
{T}huillard M. \& {Fraix-Burnet} D. 2009, {P}hylogenetic {A}pplications of the
  {M}inimum {C}ontradiction {A}pproach on {C}ontinuous {C}haracters, \emph{{E}volutionary {B}ioinformatics}, 5, 33

\bibitem[{Thuillard \& Fraix-Burnet(2015)}]{TF15}
Thuillard M. \& Fraix-Burnet D. 2015, Phylogenetic trees and networks reduce to
  phylogenies on binary states: Does it furnish an explanation to the
  robustness of phylogenetic trees against lateral transfers?, \emph{Evolutionary Bioinformatics}, 11, 213

\bibitem[{Williams \& Moret(2003)}]{Williams2003}
Williams T.~L. \& Moret B.~M. 2003, An investigation of phylogenetic likelihood methods, in \emph{Third IEEE Symposium on Bioinformatics and Bioengineering}, pp. 79--86

\end{thebibliography}
\end{document}